\documentclass[twocolumn]{aastex7}

\begin{document}

\title{A Search for Wide-orbit Planets Around M-dwarfs using Deep MIRI 15-\micron{} Images}

\author[0009-0006-7226-711X]{Yihan Li}
\affiliation{Department of Astronomy, Peking University, Beijing 100871, People’s Republic of China}
\email[show]{lyh2022@stu.pku.edu.cn}

\author[0000-0003-2969-6040]{Yifan Zhou}
\affiliation{Department of Astronomy, University of Virginia, 530 McCormick Rd., Charlottesville, VA 22904, USA}
\email[show]{yzhou@virginia.edu}

\author[0000-0001-5831-9530]{Rachel Bowens-Rubin}
\affiliation{Department of Astronomy, University of Michigan, Ann Arbor, MI 48109, USA}
\affiliation{Eureka Scientific Inc., 2542 Delmar Ave., Suite 100, Oakland, CA 94602, USA}
\email{rbowru@umich.edu}

\author[0000-0002-9521-9798]{Mary Anne Limbach}
\affiliation{Department of Astronomy, University of Michigan, Ann Arbor, MI 48109, USA}
\email{mlimbach@umich.edu}

\author[0000-0001-8274-6639]{Hannah Diamond-Lowe}
\affiliation{Space Telescope Science Institute, 3700 San Martin Drive, Baltimore, MD 21218, USA}
\email{hdiamondlowe@stsci.edu}

\author[0009-0009-1656-3020]{Cassidy E. Walker}
\affiliation{Department of Physics and Astronomy, Michigan State University, East Lansing, MI 48824, USA}
\email{cwalk@msu.edu}

\author[0000-0002-7352-7941]{Kevin B. Stevenson}
\affiliation{Johns Hopkins APL, 11100 Johns Hopkins Rd., Laurel, MD 20723, USA}
\email{kevin.stevenson@jhuapl.edu}

\author[0000-0001-7246-5438]{Andrew Vanderburg}
\affiliation{Center for Astrophysics \textbar Harvard \& Smithsonian, 60 Garden Street, Cambridge, MA 02138, USA}
\email{avanderburg@cfa.harvard.edu}

\author[0000-0002-1652-420X]{Giovanni Strampelli}
\affiliation{Space Telescope Science Institute, 3700 San Martin Drive, Baltimore, MD 21218, USA}
\email{gstrampelli@stsci.edu}

\author[0000-0002-7154-6065]{Gregory J. Herczeg}
\affiliation{Kavli Institute for Astronomy and Astrophysics, Peking University, Beijing 100871, People’s Republic of China}
\affiliation{Department of Astronomy, Peking University, Beijing 100871, People’s Republic of China}
\email{gherczeg1@gmail.com}

\begin{abstract}

Wide-orbit ($>$10 AU) gas giant planets shape the architecture of planetary systems, yet their occurrence rate remains poorly constrained. JWST has obtained the deepest mid-infrared images of nearby stars to date through substantial MIRI time-series observations of transiting planets, providing sensitive probes for wide-orbit companions. Here we leverage 15 micron observations from four programs targeting ten M-dwarf systems to search for such planets. By applying reference differential imaging for precise PSF subtraction, we achieve a 5$\sigma$ contrast of $8.9 \times 10^{-4} - 6.2 \times 10^{-3}$ (sensitivity in apparent magnitude of 14.8-15.8 mag) at a separation of 1" and $1.2 -9.1 \times 10^{-4}$ (16.5-17.9 mag) at separations $\gtrsim$3". The sensitivity is converted to planet detection probability for each system as a function of planet mass versus semimajor axis. Assuming solar metallicity and a clear atmosphere, we are sensitive to Jupiter-sized planets with an effective temperature of ${\sim}$233 K at separations beyond 20 AU in systems at 12.5 pc.
Additionally, we catalog the nearby sources and estimate their possible impact on future observations assuming they are background sources. Our results demonstrate that archival MIRI time-series imaging data is a powerful window into the population of wide-orbit gas giants around M-dwarfs.

\end{abstract}


\section{Introduction}

The James Webb Space Telescope (JWST) is a powerful planet imager. Direct imaging with JWST enables detection of cold, old, and faint exoplanets beyond the reach of previous facilities. JWST has directly imaged planets as cold as $270^\circ$K \citep{Matthews2024,Gagliuffi2025,Matthews2026,Sanghi2026} and discovered its first exoplanet TWA 7b with only 0.3 Jupiter masses through direct imaging \citep{Lagrange2025}. The JWST Mid-Infrared Instrument (MIRI) imaging can detect planets with the same temperature, mass, age, and orbital separations as Saturn and Jupiter \citep{BowensRubin2025}. The telescope has successfully imaged planets around stars with diverse stellar types \citep{Boccaletti2024,Franson2024} and found planet candidates around the nearest stellar system, Alpha Cen \citep{Beichman2025,Sanghi2025}.

MIRI has revolutionized the characterization of exoplanet thermal emission. MIRI's exceptional sensitivity in the mid-infrared wavelength range has enabled detailed atmospheric studies of planets across a wide range of temperatures and compositions \citep[e.g.,][]{Powell2024,Valentine2024}. MIRI time series observations (TSO) at 15 micron have been allocated over 300 hours of JWST General Observer time for exoplanet characterization \citep[e.g.,][]{August2025,Fortune2025,Allen2025,Meier2025, Gillon2025}. An additional 500 hours have also been allocated to the Rocky Worlds Director's Discretionary Time (DDT) program \citep{Redfield2024}. In particular, MIRI TSO have successfully captured thermal emission from temperate transiting rocky planets, which paves the way for identification and characterization of atmospheres on terrestrial exoplanets \citep{Greene2023, Zieba2023}. 

\begin{figure}
            \centering
            \includegraphics[width=1\linewidth]{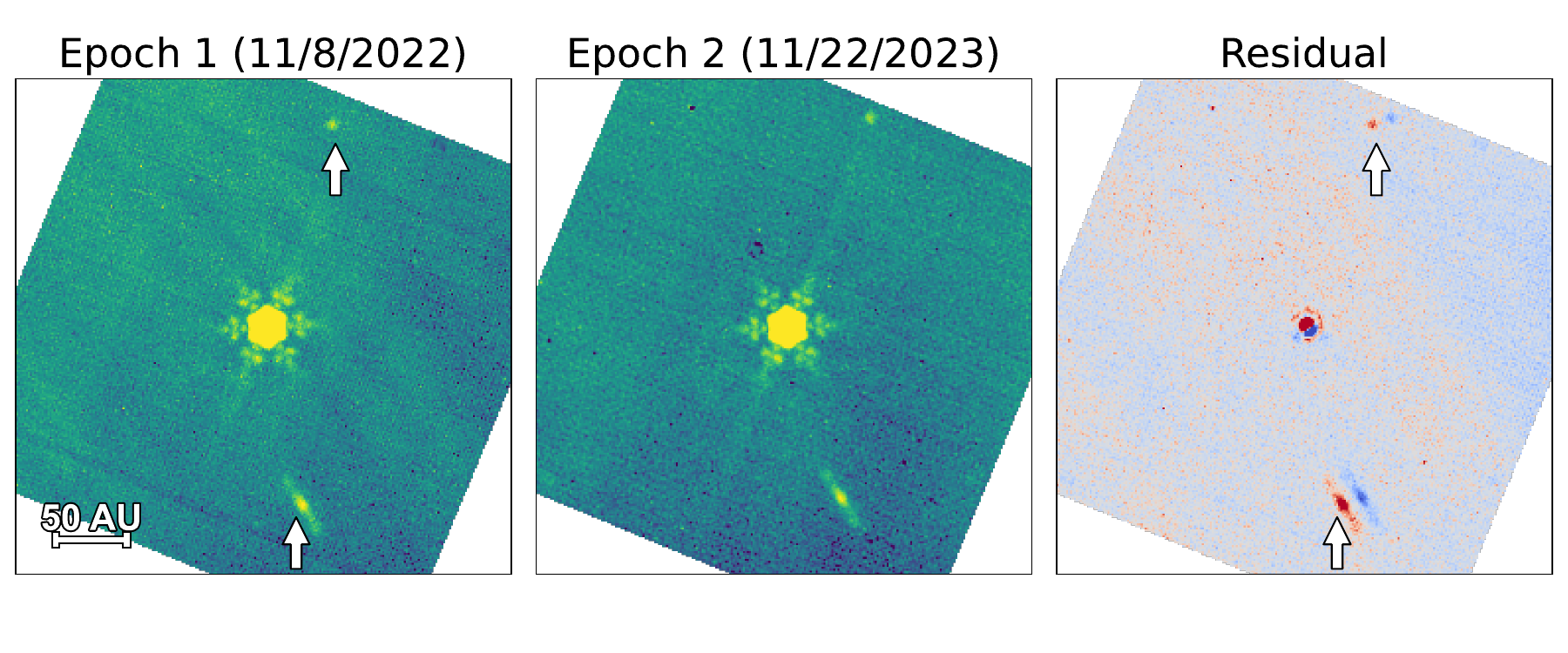}
            \caption{Examples of \texttt{cal.fits} images of TRAPPIST-1 observed in two epochs. The left panel shows time-series data obtained in 2022 (GTO-1177, Observation 7), and the middle panel shows a similar observation taken in 2023 (GO-3077, Observation 1). The right panel presents the residual image between these two epochs, with the target star aligned. There are two background objects on the image, denoted by arrows. 4. North is up and east is to the left.
            }
            \label{fig:calfits}
\end{figure}

Time-series observations can be stacked to produce deep mid-infrared images as a byproduct of their primary science goals (Figure \ref{fig:calfits}). These datasets capture wide fields of view and deep exposures \citep[][]{Greene2023, Zieba2023}. The exposure times are typically several hours and reach as high as 59 hours in phase curve observations \citep{Gillon2025}. The technique of repurposing archival imaging data for serendipitous discovery has proven successful in Solar System astronomy: a recent study reported the detections asteroids as small as 10 meters in diameter from JWST F1500W TSO data \citep{Burdanov2024}.
Similarly, the differential-imaging technique, which is a common and widely-employed technique in the high-contrast imaging community, enables searches for wide-orbit gas-giant planets in repurposed time-series observations that were originally obtained to characterize a transiting planet.

One of the priorities of the exoplanet community is studying rocky planets around M-dwarfs, as evidenced by Rocky Worlds Director's Discretionary Time program that will use 500 hours of JWST time and 250 HST orbits to survey nearby M dwarf rocky planets for the presence of atmospheres \citep{Redfield2024}. M-dwarf stars may be compelling targets for direct imaging searches of wide-orbit planets. Recent studies reveal a positive correlation between inner super Earths and outer gas giants around metal-rich, Sun-like stars \citep{Bryan2024}. This architectural relationship suggests that gas giants play a critical role in shaping planetary system formation and dynamics. \citet{Bryan2025} recently expanded their occurrence analysis for M-dwarf hosts. They found no correlations between inner super-Earths and outer gas giants at separations up to 10 au. 
While \citet{Bowler2016} suggested hints of higher gas giant occurrence rates around more massive stars compared to M-dwarfs, this trend remains statistically uncertain at wide orbital separations. Radial velocity surveys have probed gas giant occurrence around M dwarfs out to periods of $10^4$ days \citep{Pass2023}, but the occurrence rate beyond $\sim$10 au remains an open question.
Direct imaging can address this gap. Radial velocity surveys face significant challenges around M-dwarfs due to strong stellar activity \citep{Carmona2023} and the long time baselines required to detect wide-orbit companions.  Additionally, the geometric probability of transit detection for planets beyond 10 AU is less than 0.03 percent. Roman microlensing observations may detect giant planets in this parameter space \citep[e.g.,][]{Penny_2019}, but constraints from these observations are still years away. Direct imaging in the mid-infrared thus provides the most effective method to discover and characterize wide-orbit gas giants around nearby M-dwarfs. 

In this work, we present the first performance measurement of JWST/MIRI time-series data at 15 $\micron{}$ in planet direct imaging. In Section~2, we describe the archival data and reduction process. Section~3 details the method used to analyze the imaging data. Section~4 presents the performance of time-series data in both imaging sensitivity and planet detection probability. In Section~5, we compare our results with previous JWST/NIRCam and MIRI imaging performances, investigate the relationship between contrast and exposure time, and discuss the implications for M-dwarf planetary system architectures. We summarize our findings in Section~6.

\section{Data Source and Initial Processing}

The data used in this study were obtained with JWST/MIRI in time-series observation (TSO) mode in the F1500W filter. They were downloaded from the Mikulski Archive for Space Telescopes (MAST) at the Space Telescope Science Institute. The specific observations can be accessed via \dataset[doi:10.17909/e478-3553]{https://doi.org/10.17909/e478-3553}.
The dataset targeted ten M dwarfs across four JWST programs: GTO-1177, GO-2304, GO-3077, and GO-3730.
Eight out ten targets were observed during multiple visits. Each visit comprises one JWST exposure\footnote{The definition of JWST exposure is described in the \href{https://jwst-docs.stsci.edu/accessing-jwst-data/jwst-science-data-overview/jwst-time-definitions}{JWST user documentation}.}.
In total, the data set comprises 33 exposures, with an average of three exposures per target, corresponding to approximately 15.8 hours of exposure time per target.
Table~\ref{tab:target} lists the detailed target and observation information. The listed JWST/F1500W magnitude is calculated through aperture photometry using \texttt{photutils}.

\begin{deluxetable*}{lrcrcccccc}
\tabletypesize{\scriptsize}
\tablecaption{Survey Targets and Observing Information\label{tab:target}}
\tablehead{
\colhead{Name} & \colhead{Distance} & \colhead{Total integration} & \colhead{$K$} & \colhead{F1500W} & \colhead{Spectral} & \colhead{$N$} & \colhead{$N$} & \colhead{Baseline} & \colhead{Subarray}\\ \colhead{} & \colhead{(pc)} & \colhead{(hours)} & \colhead{(mag)} & \colhead{(mag)} &\colhead{Type} & \colhead{planets}&  \colhead{exposures} & \colhead{(months)} &\colhead{}
}
\startdata
GJ 3473     & $27.31 \pm 0.02$ & 12.86 & 8.829  & 8.35 & M4.0  & 2 & 4 & 7 & SUB256\\
LHS 1140    & $14.96 \pm 0.01$ & 11.02 & 8.821  & 8.20 & M4.5  & 2 & 3 & 8 & SUB256\\
LHS 1478    & $18.21 \pm 0.01$ & 6.10  & 8.767  & 8.26 & M3.5  & 1 & 2 & 2 & SUB256\\
LTT 3780    & $22.03 \pm 0.01$ & 6.00  & 8.204  & 7.74 & M4.0  & 2 & 2 & $<$1 & SUB256\\
TOI-1468    & $24.72 \pm 0.02$ & 10.84 & 8.05   & 8.17 & M3.0  & 2 & 3 & 2 & SUB256\\
TOI-270     & $22.48 \pm 0.01$ & 16.14 & 8.251  & 7.86 & M3.0  & 3 & 4 & 1 & SUB256\\
TRAPPIST-1  & $12.47 \pm 0.01$ & 86.35 & 10.296 & 9.48 & M7.5e & 7 & 11 & 12 & FULL+BRIGHTSKY\\
HD 260655   & $9.998 \pm 0.002$ & 6.73 & 5.862  & 5.54 & M0.0  & 2 & 2 & $<$1 & SUB64\\
L 98-59     & $10.608 \pm 0.002$ & 3.34 & 7.101 & 6.71 & M3.0  & 5 & 1 & -- & SUB128\\
GJ 357      & $9.436 \pm 0.002$ & 4.03 & 6.475 & 6.06 & M2.5  & 3 & 1 & -- & SUB64\\
\enddata
\tablecomments{The information of our targets. Distances are calculated from Gaia Data Release 3 \citep{Gaiamission, Gaiacontents}. Total integration time includes only on-detector exposure time, excluding instrument reset periods.
K-band magnitude and spectral type are obtained from SIMBAD. 
The ``N planets" column represents the number of known, transiting planets based on NASA Exoplanet Archive.
The Baseline column lists the longest time baseline of multi-epoch observations of the target.}
\end{deluxetable*}

The first three programs, GTO-1177, GO-2304, and GO-3077, targeted TRAPPIST-1, an M-dwarf hosting seven transiting terrestrial planets \citep{2017Natur.542..456G, Ducrot2025}. The last program, GO-3730, targeted nine M-dwarfs hosting transiting rocky planets \citep[e.g.,][]{August2025}. The F1500W ($\lambda_0=15.0$~\micron{}, $\Delta\lambda=2.92$~\micron{}, FWHM=0.488") imaging time series were initially acquired to characterize the secondary eclipses of the transiting terrestrial planets. The long exposures provide sensitivity enabling the direct detection of wide-orbit giant planets. 

We downloaded the \texttt{jwst} Stage 1 pipeline products (\texttt{rate} files) from the MAST archive. The \texttt{rate} files were used instead of the \texttt{rateints} files, because the point spread function (PSF) is extremely stable between integrations of the same exposure. Therefore, using individual integrations provided in \texttt{rateints} files does not enhance the diversity of reference PSF or improve PSF subtraction but substantially increases the computational cost \citep{2024AJ....168...51K}.
The \texttt{rate} files were then processed with the \texttt{jwst} Stage 2 pipeline (version 1.17.1). In this step, WCS is assigned, the flat field is calibrated, and the flux units are converted to MJy/sr. We unified the data dimensions by cropping all images (except for three targets listed below) to $252\times252\ \mathrm{pixel}^2$ centered on the target (field of view: 27.8" $\times$ 27.8", pixel size: 0.11"). Three targets are observed with even smaller subarrays: L 98-59 uses SUB128, and target HD 260655 and GJ 357 use SUB64. Therefore, we cropped these images to $40\times40\ \mathrm{pixel}^2$ (field of view: 4.4" $\times$ 4.4").

To prevent spurious pixels from affecting PSF subtraction analysis, we identified and corrected three types of ``bad pixels". The first type is pixels flagged as “DO NOT USE” by the \texttt{jwst} pipeline. The second type is spurious background pixels missed by the pipeline. To identify them, we applied a sliding window (10 pixels $\times$ 10 pixels) and flagged isolated pixels with values exceeding the local median by $5\sigma$. Regions within 25 pixels of the central star were excluded because the large flux gradient makes median and standard deviation in a square window unreliable. The third type is bright isolated pixels lying within 25 pixels of the stellar centroid. These anomalous pixels appeared in small numbers, with typically one or two detected per exposure.
We identified these pixels using the same $5\sigma$ criterion but with two annuli windows (r=13-19\,pixels, r=19-25 \,pixels). We replaced all detected pixels by bilinear interpolation from neighboring pixels.

\section{PSF Modeling and Subtraction}

\subsection{KLIP}

PSF subtraction is performed via KLIP, an algorithm based on Principal Component Analysis (PCA) \citep{Amara&Quanz2012, 2012ApJ...755L..28S}. PCA identifies the dominant stellar PSF patterns across images, and uses these patterns to construct a PSF model. We employ reference differential imaging (RDI) to construct and subtract the host star PSFs. For each set of target images, a dedicated library of reference images is assembled, which consists of images from other stars with similar PSF characteristics. The reference library serves as a template for modeling the target star PSF. We implement this approach using \texttt{pyKLIP} \citep{Wang2015}.

\subsection{PSF Library} \label{sec:library}

We first assemble a PSF library to perform RDI on each target star. A key challenge is that the PSF shape depends on detector position. For our data, the target position on the detector is determined by the choice of subarray, with each subarray placing targets at a fixed location. Therefore, we categorize the PSF library by subarray type, using only images taken with the same subarray as mutual references. Table~\ref{tab:target} lists the observation subarray used for each target star. However, two exceptions apply. First, TRAPPIST-1 was observed with the FULL (GTO-1177 and GO-2304) and BRIGHTSKY (GO-3077) subarrays, neither of which is shared by any other target in our study. Consequently, we use images from the other nine targets as references for TRAPPIST-1 despite the subarray mismatch.
Second, three bright sources were observed with small subarrays (SUB128 and SUB64), which collectively provide only ten reference images.
To ensure a sufficiently large PSF library, we include all images targeting the other seven stars as references, regardless of subarray.

We align the target and reference images by PSF centers. The PSF centers are determined by fitting model PSFs. To account for the detector's geometric distortion and the resulting spatial variations in the PSF shape, we create a grid of 49 model PSFs using \texttt{stpsf} distributed across the full detector.
Next, a \texttt{photutils} PSF model class, \texttt{GriddedPSFModel}, is created using the model PSF grid. All cropped images are extended to the size of the full detector according to their original subarray configurations, so that the target’s coordinates in the extended image match its actual location on the detector. We use \texttt{photutils} to interpolate four PSFs in the grid that are closest to the target position to get a model PSF. We find the best-fitting scaling and centroid coordinates using a nonlinear least-squares approach \citep{larry_bradley_2025_14606896}. The images in the PSF library are then aligned to a common central coordinate.

\subsection{KLIP Parameter Optimization}\label{sec:optimization}

Optimizing the KLIP parameters is essential, as they significantly affect PSF subtraction performance. During KLIP modeling, images are divided into multiple annuli and position-angle (PA) sectors. KLIP is performed on these image subregions independently. The number of principal components (KL modes) controls the aggressiveness of the PSF subtraction. Increasing both the number of subregions and principal components allows for more precise modeling of the stellar PSF and more thorough suppression of starlight. However, this also increases the risk of oversubtraction and potentially removing planetary signals \citep{Pueyo2016}. Here we follow the method described in \citet{AdamsRedai2023} to explore the optimal KLIP parameters.

The injection and recovery of model PSFs is a key step for evaluating the PSF subtraction performance \citep{Pueyo2016}. In our case, we generate synthetic planet PSFs by interpolating the model PSF grid produced by \texttt{stpsf} at the host star position (also see Section~\ref{sec:library}). 
These synthetic PSFs, with fixed total flux approximately 20 times the local noise level, are injected into the \texttt{cal.fits} images at separations of 5, 20, 60, and 100 pixels. To account for azimuthal variations, we place four PSFs at each separation, spaced 90 degrees apart in position angle. 
For the three targets using smaller subarrays, we inject only one PSF at each of three separations (5, 10, and 15 pixels), with the three PSFs separated by 90 degrees in position angle.
The modified images are then processed with KLIP using identical parameters as the science data. After PSF subtraction, we measure the recovered flux of each injected signal within a circular aperture of diameter 4.436 pixels, corresponding to the PSF FWHM for F1500W from the JWST documentation\footnote{\url{https://jwst-docs.stsci.edu/jwst-mid-infrared-instrument/miri-performance/miri-point-spread-functions}}.

We evaluate PSF subtraction performance using the 5$\sigma$ contrast curve. It is calculated as follows. Following \citet{Mawet2014}, we place rings of apertures (diameter 4.436 pixels) at separations of 3–100 pixels and compute the raw 5$\sigma$ noise with small-sample statistics corrections. The throughput correction factor is the mean recovered/injected PSF flux ratio across the four PSFs at each separation. We linearly interpolate the throughput curve from injected PSF separations (5, 20, 60, and 100 pixels) and extrapolate inward to 3–5 pixels. We derive the calibrated 5$\sigma$ noise by dividing the raw 5$\sigma$ noise by the throughput curve. To convert the 5$\sigma$ noise to contrast units, we measure the target star brightness via aperture photometry with an aperture identical to that used for the planet. An annulus of 40–50 pixels (15–20 pixels for targets with small fields of view) defines the background region, which is subtracted from the raw star flux. The final contrast curve, in planet-to-star flux ratio units, is obtained by dividing the calibrated 5$\sigma$ noise curve by the estimated star flux.

We select the optimal value for each KLIP parameter set using image quality metrics proposed by \citet{AdamsRedai2023}: (1) Contrast: the calibrated $5\sigma$ contrast curve evaluated at separations of 5, 20, 60, and 100 pixels; (2) Peak S/N: the peak signal-to-noise ratio of the recovered model PSF; (3) Neighbor Quality: the Peak S/N smoothed over the KL modes–annuli parameter space with a Gaussian kernel; and (4) False Positive Fraction: the fraction of pixels in the PSF-subtracted image, within the inner and outer working angles, that exceed the local $5\sigma$ noise level.

We normalize each metric following the procedure described in \citep{AdamsRedai2023}, scaling values so that the best-performing parameter set has value 1 and the worst has value 0. For Peak S/N and Neighbour Quality, the best parameter set gives the highest S/N of a recovered PSF injection. For Contrast, the best set produces the lowest $5\sigma$ contrast at injection separations. For False Positive Fraction, the best set minimizes the number of false positive pixels. Finally, we compute the average of the four metrics: Peak S/N, Neighbor Quality, Contrast, and False Positive Fraction to evaluate the parameters. Since the False Positive Fraction is nearly uniform across the parameter space and contributes negligibly to distinguishing between parameter sets, we exclude it from the final evaluation. Instead, we take the average of the remaining three metrics as the standard for selecting the optimal parameter set.

\begin{figure}
  \centering
  \includegraphics[width=1\linewidth]{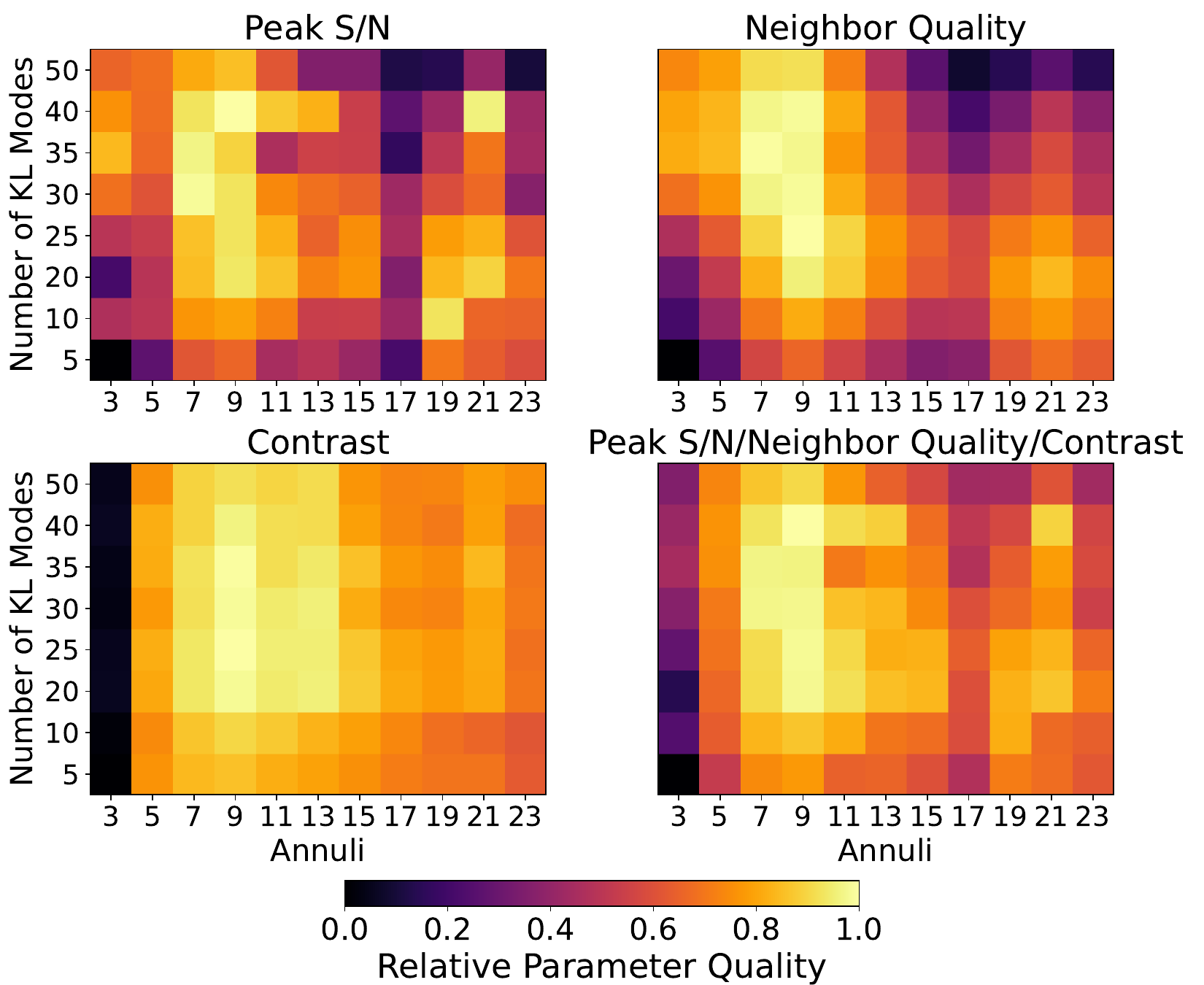}
  \caption{Map of image metrics across numbers of KLIP \texttt{basis} and \texttt{annuli} parameter space for TRAPPIST-1. 
  The lower right show the quality of parameters measured by the combination of these metrics. 
  Each pixel's color represents the relative quality of that parameter combination, with 1 being the best and 0 being the worst. The False Positive Fraction metric is nearly uniform across parameter space and has minimal effect on the combined metrics. We therefore selected optimal parameters based on the combined Peak S/N, Neighbor Quality, and Contrast metrics. }
  \label{fig:metrics}
\end{figure}

We explore the number of KL modes and annuli parameter space and determine the optimal number of KL modes and annuli parameters for each target using the method described above. Figure~\ref{fig:metrics} shows the image metric maps across parameter space for TRAPPIST-1. For the seven targets with larger fields of view, the optimal number of KL modes falls between 10 and 40, with annuli values ranging from 9 to 23. For the other three bright sources, the optimal number of KL modes lies between 5 and 25, and the corresponding annuli range from 4 to 6. Due to the small image size of the three bright sources, we explored relatively small annuli values (1 to 8) to ensure there are enough pixels inside each annulus. 

\begin{deluxetable}{lcc}
\tabletypesize{\small}
\tablecaption{Optimal KLIP Paramters\label{tab:params}}
\tablehead{
\colhead{Target} & \colhead{Number of} & \colhead{Number of}\\
\colhead{}       & \colhead{KL modes}   & \colhead{Annuli}
}
\startdata
GJ 3473 & 25 & 9 \\
LHS 1140 & 25 & 23 \\
LHS 1478 & 20 & 21 \\
LTT 3780 & 40 & 13 \\
TOI-1468 & 25 & 9 \\
TOI-270 & 10 & 19 \\
TRAPPIST-1 & 40 & 9 \\
HD 260655 & 15 & 4 \\
L 98-59 & 5 & 6 \\
GJ 357 & 25 & 4 \\
\enddata
\end{deluxetable}

\subsection{Image Analysis} \label{sec:image analysis}

\begin{figure}
            \centering
            \includegraphics[width=1\linewidth]{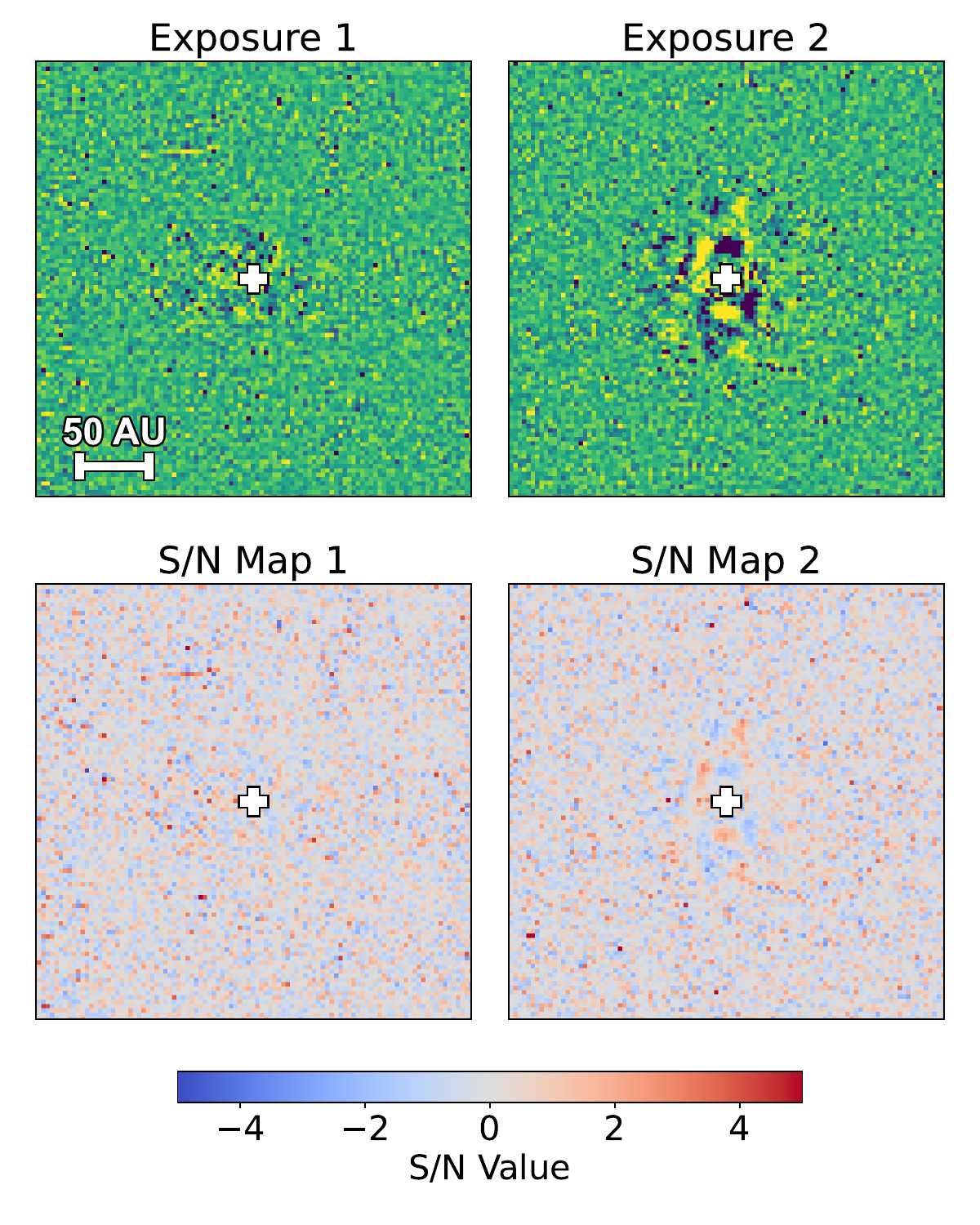}
            \caption{PSF-subtracted images of two exposures targeting GJ 3473 and the corresponding S/N maps. The upper left image shows a clean PSF subtraction while the upper right image presents apparent residuals near the central star. The peak S/N of these residual patterns is $\sim$2.8 (lower right panel). North is up and east is to the left.}
            \label{fig:exposures}
\end{figure}

We perform PSF subtraction on each target using the optimal KLIP parameters listed in Table~\ref{tab:params} and obtain the PSF-subtracted images. Upon examining these images, we identify marginally significant residuals (peak S/N 2-3) within 15 pixels of the stellar center for four targets. An example of GJ 3473 is presented in Figure~\ref{fig:exposures} (upper right panel). 

To determine whether these residuals are artifacts or real signals, we conduct two diagnostic tests. First, we perform KLIP using ten different subsets of the reference image library to determine whether the residuals are artifacts arising from specific reference frame combinations. The residuals persist across all reference subsets with peak S/N values consistently between 2 and 3, indicating they are not caused by particular reference PSF selections.

Second, we perform KLIP on each individual exposure of a given target separately to test whether these residuals are artifacts that originate from specific exposures. As shown in Figure \ref{fig:exposures}, all targets with strong central residuals contain one individual exposure that already displays similar residual features before stacking. Since the exact origin of these residuals remains unclear and requires further investigation, we conservatively exclude the affected exposures from our analysis and present KLIP results based only on the remaining clean exposures.

\begin{figure*}[t]
  \centering
  \includegraphics[width=\textwidth]{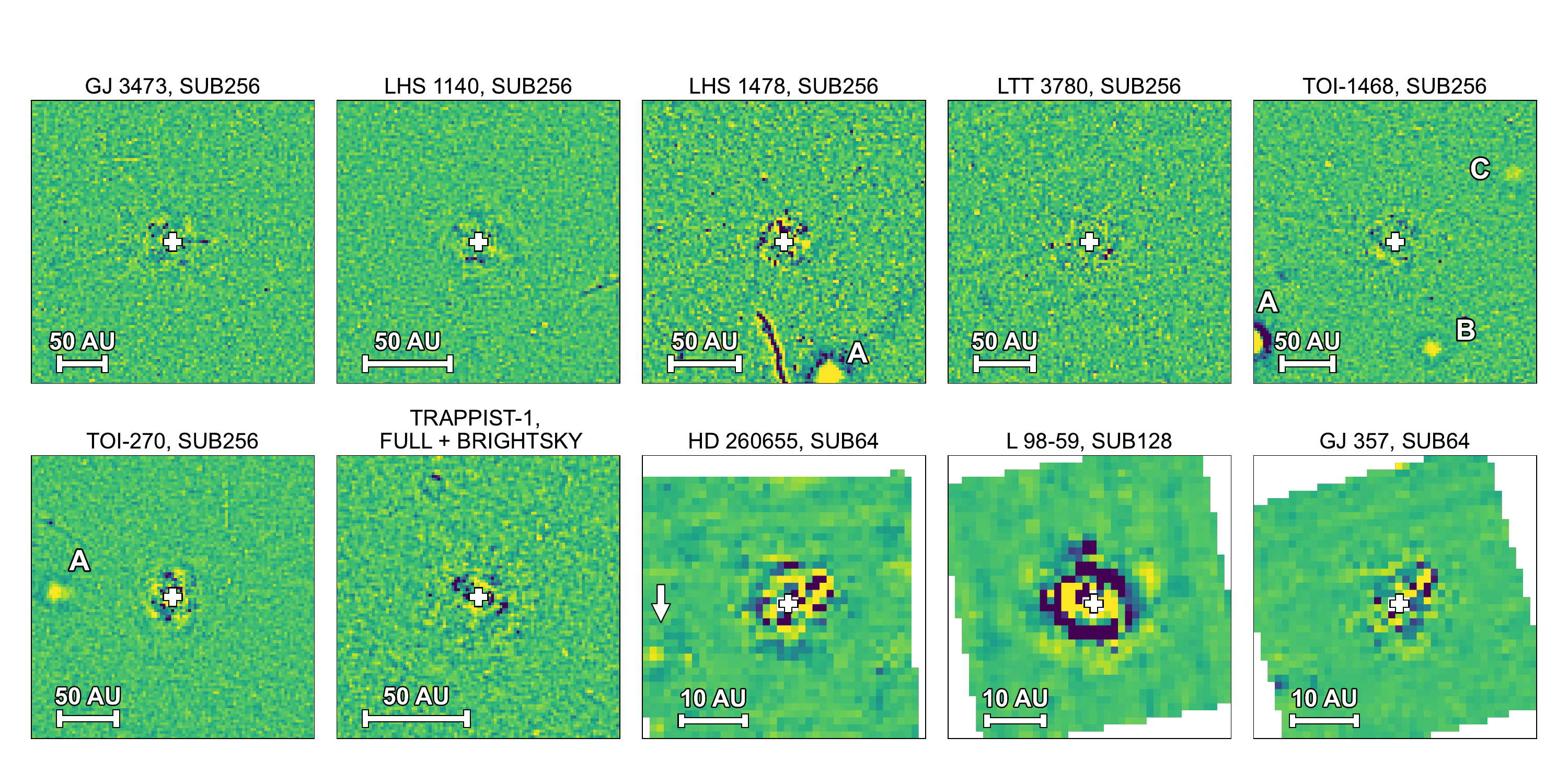}
  \caption{PSF-subtracted images of all targets. The observation subarray of each target is indicated. The first seven images are zoomed in to area of 11.0" $\times$ 11.0" for clearer view of the central regions. For the last three targets, the small size (4.4" $\times$ 4.4") is due to the choice of the observation subarray. The location of the star PSF center is denoted by the white cross. Sources around LHS 1478, TOI-1468, and TOI-270 are labeled with letters, which are likely background sources.
  The arrow near HD 260655 indicates a low-S/N identification of possible close companion. We further examined this detection in Section~\ref{sec: close residual}. North is up and east is to the left.}
  \label{fig:image}
\end{figure*}

Figure~\ref{fig:image} shows the PSF-subtracted images of all targets after removing the affected exposures. $100\times100$ pixel area ($11.0'' \times 11.0''$) highlights the region close to each central star for the first seven targets. Figure~\ref{fig:S/N} shows the corresponding S/N maps of these images.

Figure~\ref{fig:contrast} shows the $5\sigma$ contrast curves, computed as in the same way described in Section~\ref{sec:optimization}.
Additionally, we convert the 5$\sigma$ contrast curve to apparent magnitude sensitivity.
We first apply aperture correction via the \texttt{stpsf} encircled energy function to recover the total stellar flux. The total planet flux is the product of the total stellar flux and the 5$\sigma$ contrast. We convert the resulting flux to apparent magnitude with the \texttt{flux\_to\_magnitude} function from the \texttt{species} package \citep{Stolker2020}.
Figure~\ref{fig:contrast} shows the 5$\sigma$ contrast and apparent magnitude sensitivity curves for all targets.

\section{Results}

\subsection{Primary Subtraction Results} \label{sec: subtraction results}

\begin{figure*}[t]
  \centering
  \includegraphics[width=\textwidth]{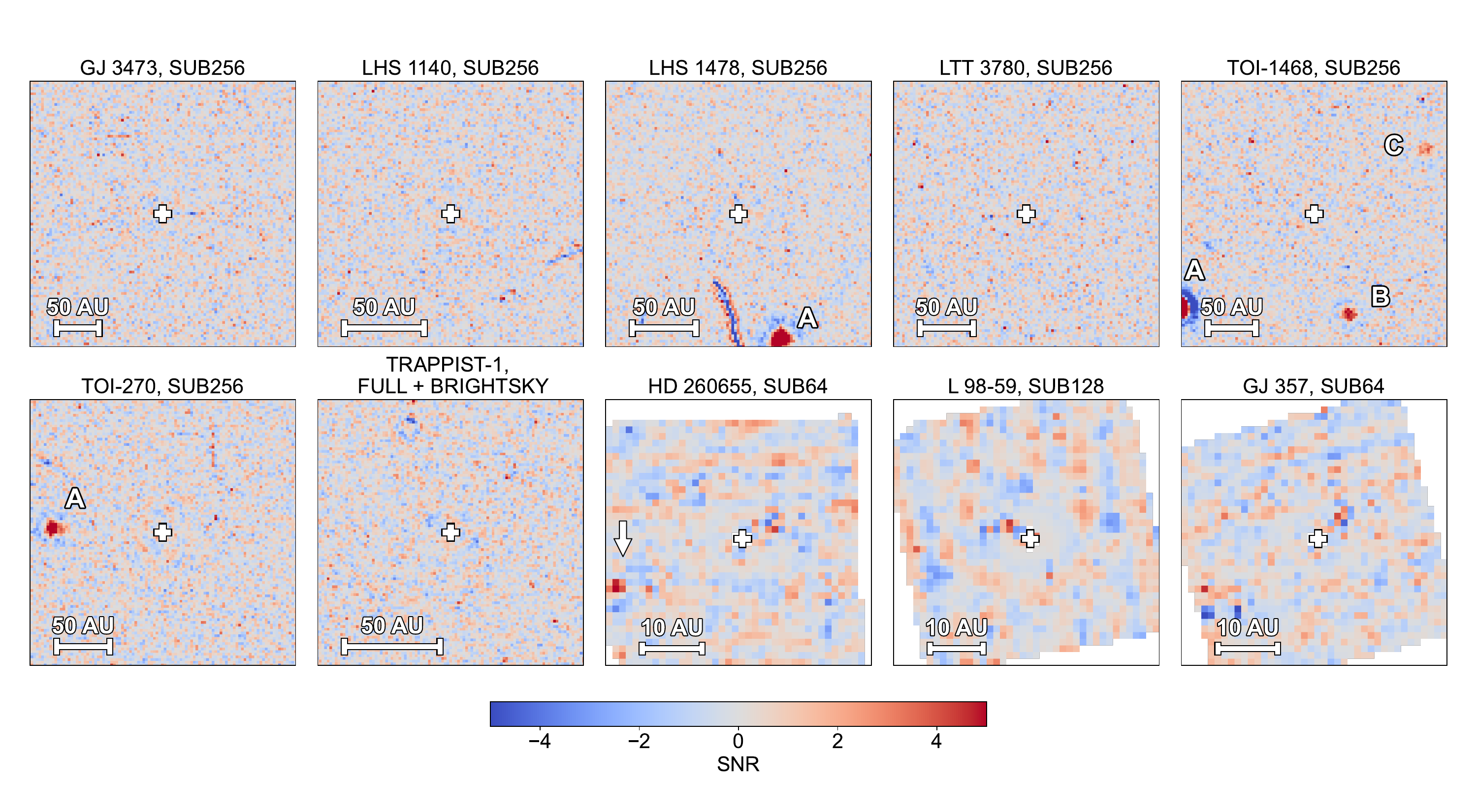}
  \caption{Signal-to-noise ratio maps of all targets. Figure annotations follow the same convention as Figure~\ref{fig:image}.}
  \label{fig:S/N}
\end{figure*}

Figure \ref{fig:image} and Figure \ref{fig:S/N} present the PSF-subtracted images and their corresponding S/N maps. Three of our targets show point sources within $5.5''$ of the central star. 
Source LHS 1478-A has S/N = 24.1. Source A, B, and C around TOI-1468 has S/N = 47.6, 11.2, and 6.7. Source TOI-270-A has S/N = 14.1.

Additionally, HD 260655 shows a point-source-like signal with S/N of 4.86 (denoted by an arrow) within 2". We present the 1D profile in Section~\ref{sec:close sources}. 
LHS 1478 and LHS 1140 both show line-shaped detector artifacts located to the south and west, respectively.

\begin{figure}
            \centering
            \includegraphics[width=1\linewidth]{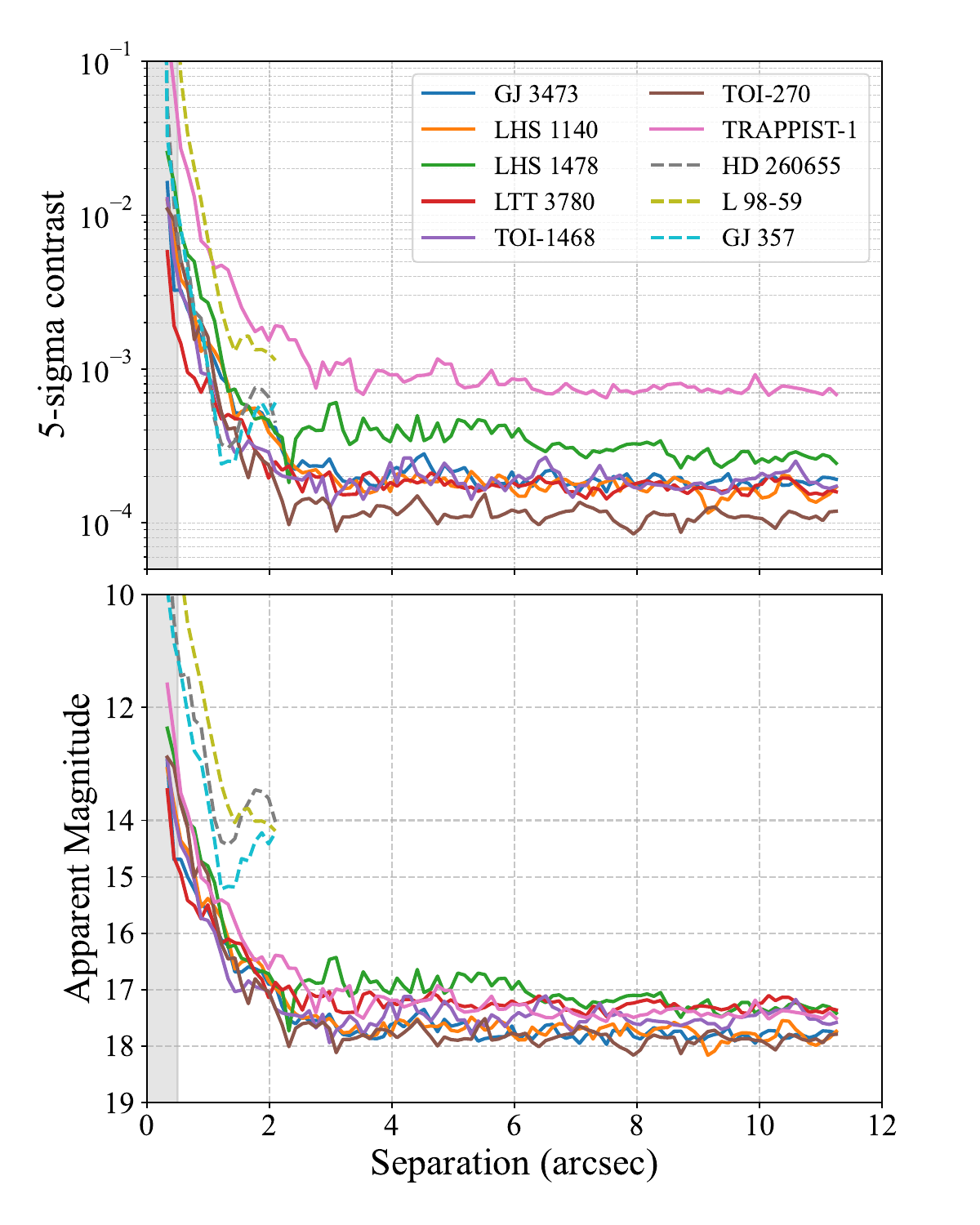}
            \caption{Survey sensitivity in 5$\sigma$ contrast (upper panel) and apparent magnitude (lower panel). The contrast is in planet flux/star flux unit, while the apparent magnitude corresponds to sensitivity. The three bright targets HD 260655, L 98-59, and GJ 357 are observed with smaller subarrays, thus limiting the curves to $\sim$2". The shaded area within 0.5" indicates the region there the throughput curve is extrapolated (see Section~\ref{sec:optimization}). The sensitivity in contrast differs from target to target, due to differences of central star brightness and exposure time.}
            \label{fig:contrast}
\end{figure}

The 5$\sigma$ contrast curves and sensitivities in apparent magnitude are shown in Figure \ref{fig:contrast}. The contrast is listed in planet flux/star flux unit, while the apparent magnitude corresponds to 5$\sigma$ sensitivity. We achieve a median 5$\sigma$ contrast of $1.1 \times 10^{-2}$ (median sensitivity in apparent magnitude of 13.7 mag) at a separation of 0.5", and $2.1 \times 10^{-4}$ (17.4 mag) at separations $\gtrsim 3$". The contrast curves become deeper as the separation increases within the first 2". Beyond that, the curves become background-limited and remain roughly constant.

Differences of 5$\sigma$ contrast and sensitivity between targets are mainly due to different luminosity and exposure time of each source. With the same exposure time, 5$\sigma$ noises are of the same order in the background-limited region. As a result, fainter targets tend to have a worse contrast. 
Longer exposure times yield greater apparent magnitude sensitivity, with a $\sim$1 magnitude range across our targets. However, TRAPPIST-1 does not follow this trend despite its substantially longer integration time. 
Its proximity to the ecliptic plane may explain this: the elevated zodiacal background increases noise and offsets the gain from longer integration time. Targets HD 260655, L 98-59, and GJ 357 are much brighter, so they were observed with SUB128 and SUB64 subarrays to avoid saturation. Consequently, their contrast curves are limited to separations within $\sim $ 2". 
In addition, as described in Section~\ref{sec:library}, we include images using different subarrays as references. Their different positions on the detector hindered our ability to capture the detailed structure of the star PSF, leading to poorer PSF subtraction. As a result, the contrast curves for these bright sources are not as deep, as shown in Figure~\ref{fig:contrast}.

\subsection{Detection Sensitivity} \label{sec:sensitivity}

We use the observed contrast curves to evaluate planet detection sensitivity as a function of planet mass and semimajor axis following the approach developed by \citet{Bonavita2012} and demonstrated for JWST images by \citet{bogat2025probing}.Substellar evolutionary models, \texttt{BEX-petitCODE} \citep{Linder2019} and \texttt{ATMO-CEQ} \citep{Phillips2020}, are adopted to translate apparent magnitude limits into mass limits at specific ages. We use Monte Carlo calculations to convert semimajor axes into projected angular separations, allowing us to compute detection probabilities across a range of orbital configurations.

To implement this approach, we create a mass-semimajor axis grid ranging from 3 to 20 $\mathrm{M_{Jup}}$ and 1 to 500 AU, sampled with 100 mass values and 100 semimajor axis values uniformly spaced in logarithmic scale.
For each planet mass in our grid, we estimate a planet's F1500W absolute magnitude by interpolating evolutionary model grids. We then convert these absolute magnitudes to apparent magnitudes using target distances from Gaia Data Release 3.
For masses below 2 $\mathrm{M}_{\mathrm{Jup}}$, we use the \texttt{BEX-petitCODE} model. We use the solar metallicity and cloudless assumptions as our nominal case and explore the impact of different metallicity and cloud assumptions below. For masses above 2 $\mathrm{M}_{\mathrm{Jup}}$, we use the \texttt{ATMO-CEQ} chemically equilibrium model. This combination follows the approach of \citet{bogat2025probing}.

Four of our targets have age constraints: TRAPPIST-1 (7.6 $\pm$ 2.2 Gyr; \citep{Burgasser2017}), LHS 1140 ($>$ 5 Gyr; \citep{Dittmann2017}), HD 260655 (2-8 Gyr; \citep{Luque2022}), and L 98-59 (4.94$\pm$0.28 Gyr; \citep{Engle2023}).
Therefore, we adopt 5 Gyr as the nominal age for targets without age constraints, and explore alternative ages 1 Gyr and 10 Gyr.

For each semimajor axis in our grid, we calculate a projected separation via Monte Carlo sampling. Following \citet{bogat2025probing}, we assume a Gaussian distribution with mean of 0 and sigma of 0.1 for orbital eccentricities, an isotropic distribution of inclination (i.e., $\cos i$ is uniformly distributed, $i$ is the inclination angle), and a uniform distribution for the argument of periapsis. To simplify the calculation, we also assume a uniform distribution of the true anomaly, which serves as a good approximation for nearly circular orbits. At each grid semimajor axis, the projected separation is calculated accordingly. A model planet is considered detectable if its apparent magnitude is brighter than the sensitivity at the projected separation. This Monte Carlo simulation is repeated 1,000 times per grid point to derive the detection probability across the parameter space. 

\begin{figure*}[t]
  \centering
  \includegraphics[width=\textwidth]{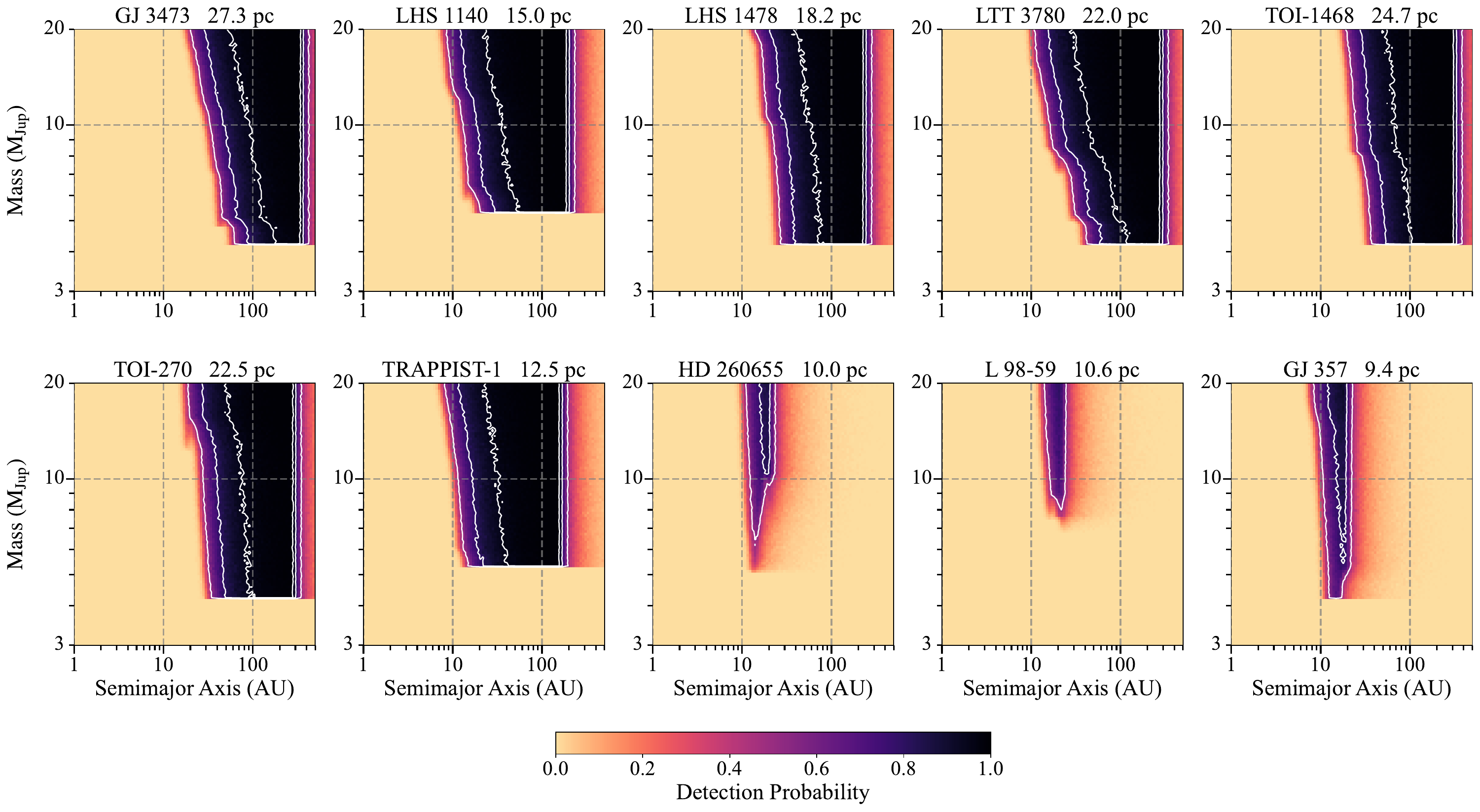}
  \caption{Detection probability map as a function of planet mass versus semimajor axis, assuming a solar metallicity and cloud-free atmosphere model. The age assumption is 5 Gyr, except for LHS 1140 (8 Gyr), TRAPPIST-1 (7.6 Gyr), and L 98-59 (4.94 Gyr). White contours show detection probabilities of 50\%, 80\%, and 95\%. The truncation below 4 $\mathrm{M}_{\mathrm{Jup}}$ is caused by the lack of evolutionary model points at $>$5 Gyr (see Section~\ref{sec:sensitivity} for details).
  The last three targets are bright and used smaller subarrays. Therefore, their maps truncate at $\sim$30 AU.}
  \label{fig:sensitivity}
\end{figure*}

Figure~\ref{fig:sensitivity} shows the detection sensitivity calculated assuming a solar metallicity and cloud-free atmospheric model. We assume that the stellar ages are 8 Gyr for LHS 1140, 7.6 Gyr for TRAPPIST-1, 4.94 Gyr for L 98-59, and 5 Gyr for the others. At 5 Gyr or older, the \texttt{BEX-petitCODE} grid does not have model less massive than 2 $\mathrm{M}_{\mathrm{Jup}}$, and the \texttt{ATMO-CEQ} grid only covers $M\gtrsim$4 $\mathrm{M_{Jup}}$. Therefore, the map shows a sharp cutoff at around 4 $\mathrm{M}_{\mathrm{Jup}}$. Our sensitivity at close separation is limited by contrast. There is a inner limit, which is an analog of the inner working angle, within which no planets below 20 $\mathrm{M}_{\mathrm{Jup}}$ can be detected. This inner limit ranges from 7 AU (represented by LHS 1140) and 20 AU (represented by GJ 3473).

Except for the three targets observed with small subarrays, we are sensitive to planets as small as 10 $\mathrm{M}_{\mathrm{Jup}}$ at 10 AU semimajor axis. The sensitivity improves with increasing semimajor axis. For solar metallicity and cloud-free atmospheres, we are sensitive to planets with Jupiter-like radius and an effective temperature of $\sim$233 K at separations beyond 20 AU in systems at 12.5 pc. Here we adopt the 1-bar reference temperature of Jupiter from \citet{Gupta2022} as a proxy for the effective temperature in our atmospheric model, as this pressure level serves as the standard anchor for Jupiter's thermal structure. For the three bright targets observed with smaller subarray, the sensitivity also drops to zero within 10 AU semimajor axis. At around 10 AU, we are sensitive to as small as 4 $\mathrm{M}_{\mathrm{Jup}}$. Beyond 30 AU, the sensitivity for these three targets is cut off due to the limited image field of view.

\begin{figure*}[t]
  \centering
  \includegraphics[width=\textwidth]{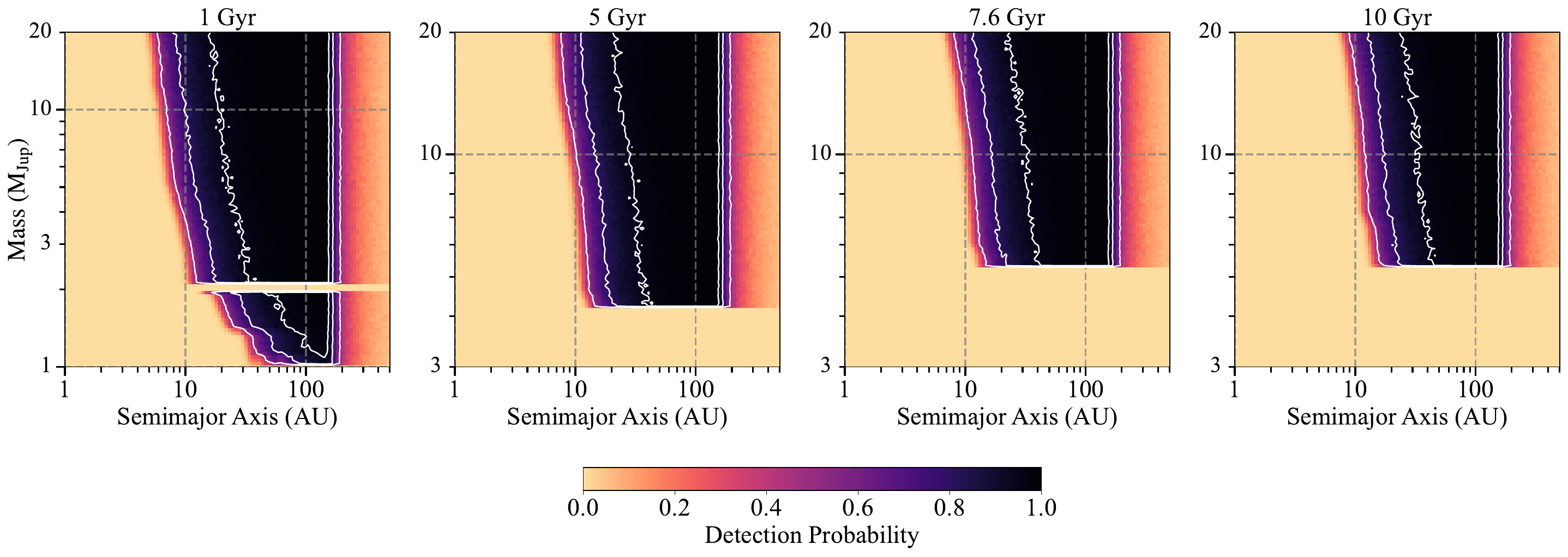}
  \caption{Detection probability maps for TRAPPIST-1, considering different ages, with solar metallicity and cloud-free atmospheres. Detectable mass limits increase with system age. The cutoff at the bottom is a artifact caused by the lack of evolutionary model points at such old age. The discontinuity in the 1 Gyr map is a result from different choice of model for below and above 2 $\mathrm{M}_{\mathrm{Jup}}$.
  }
  \label{fig:TRAPPIST}
\end{figure*}

We use TRAPPIST-1 to demonstrate how stellar age affects detection probability. The detection sensitivity worsens as the assumed age increases for cloud-free, solar metallicity models (Figure \ref{fig:TRAPPIST}). For a system at 12.5 pc with the age of 1 Gyr, we can detect a 2 $\mathrm{M}_{\mathrm{Jup}}$ planet at 20 AU with 80\% probability.

\subsection{Close Companion Sources} \label{sec:close sources}

\begin{figure*}[t]
  \centering
  \includegraphics[width=\textwidth]{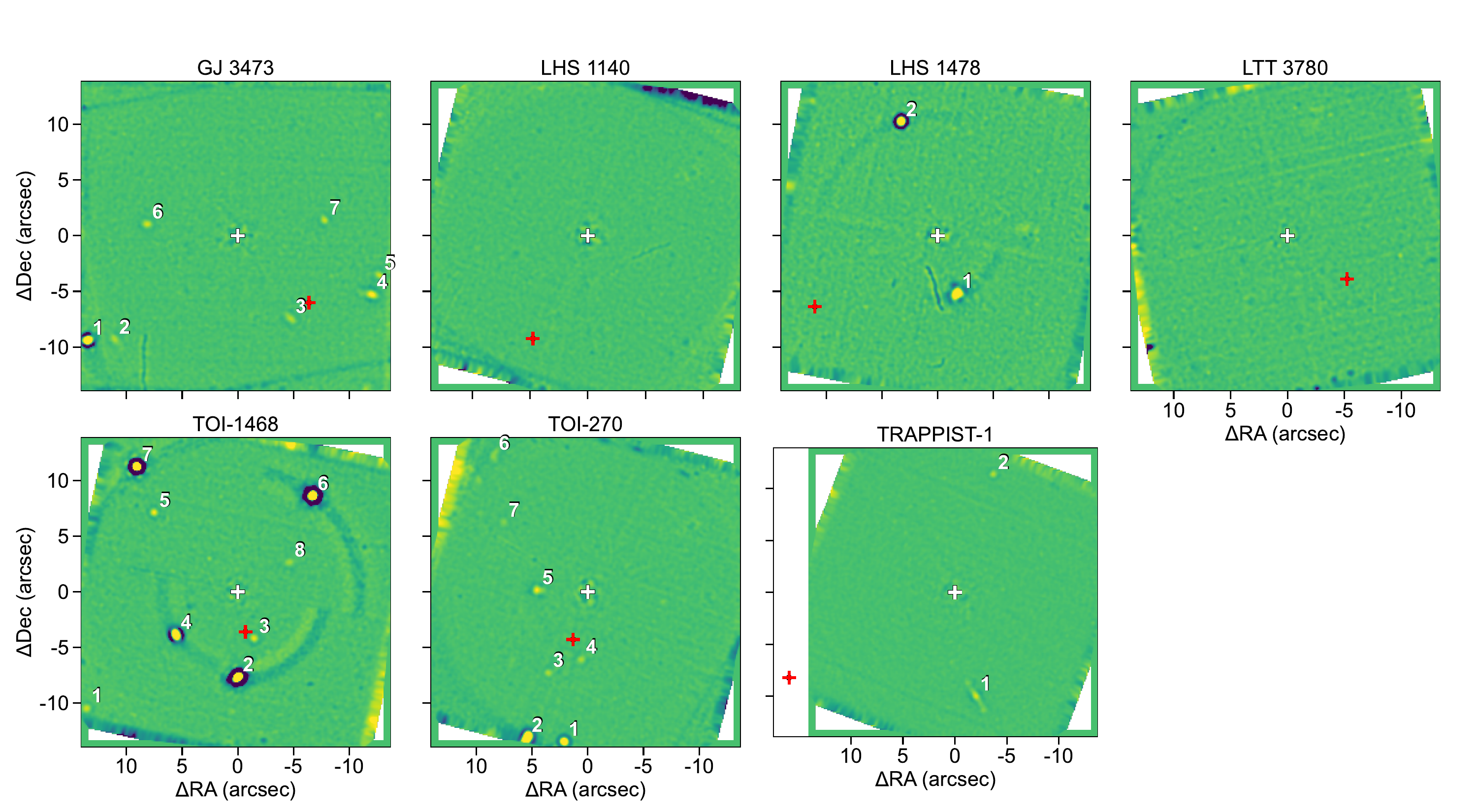}
  \caption{ 
  PSF subtracted images of all targets, convolved with a Gaussian kernel to better reveal close companion sources. The central star position is denoted by the white cross. Detected sources are labeled, and their astrometry and photometry properties are achieved in Table~\ref{tab:bkg}. The red cross on each subplot represents the position of the target star at J2040. The plots show that the target star and the companion sources (assuming background objects) could get close due to differences between their proper motions. Future observations might need to take this issue into account.}
  \label{fig:bkg}
\end{figure*}

Figure~\ref{fig:bkg} shows that for stars GJ 3473, LHS 1478, TOI-1468, TOI-270, and TRAPPIST-1, we detect nearby sources within 50 arcsec. 
We measure their astrometry and photometry using PSF forward modeling with the \texttt{pyKLIP.fm} module. The FWHM of each source is measured using \texttt{lmfit} \citep{Newville2025} to distinguish between point sources and extended objects. We estimate the potential masses of point sources assuming they are substellar companions. 
Furthermore, we predict the future relative positions of these sources with respect to the target stars under the assumption that they are background objects.

PSF forward modeling is a necessary step to correct for flux loss caused by PSF subtraction \citep{Pueyo2016}. We inject a model PSF with chosen flux at the same position as the background source in a simulated blank image matching the science image size. PSF subtraction is performed to the simulated image using the same KLIP parameters as the science image to create a forward model. This model is then fit to the companion source using a Markov Chain Monte Carlo (MCMC) algorithm. This yields the best-fitting flux scaling factor and coordinates of this source. To assess source morphology, we fit the FWHM of each source using a custom rotational 2D Gaussian model with \texttt{lmfit}. Sources whose major axis FWHM exceeds 3$\sigma$ of the expected point source FWHM (4.436 pixels) are categorized as extended sources. Table~\ref{tab:bkg} lists the relative positions to the star, detection S/N, brightness, and fitted FWHM along major and minor axes.

We estimate the masses of these companion sources assuming they are substellar companions using same evolutionary model configuration as described in Section~\ref{sec:sensitivity}. We obtain the masses from the models at ages of 1 Gyr, 5 Gyr, and 10 Gyr using the calculated brightnesses.
For 1 Gyr, the lower magnitude limit of the \texttt{ATMO-CEQ} model is 14.22 mag, while the upper limit of the \texttt{BEX-petitCODE} model is 14.84 mag. Therefore, magnitudes between these two values lie outside both models. We did not extrapolate the \texttt{ATMO-CEQ} grids in age or extrapolate the \texttt{BEX-petitCODE} grids in magnitude to keep our results robust, instead we give the upper and lower mass limit predicted by the two grids. The estimated mass for each source is listed in Table \ref{tab:bkg}.
As shown in Figure~\ref{fig:calfits}, two programs targeting TRAPPIST-1 at different epochs demonstrate that the two sources near TRAPPIST-1 are not co-moving with the star. We therefore rule them out as substellar companions and classify them as background sources in the table.

We assess the possible impact of these sources on future observations, considering the differences in proper motion between the planet host stars and background sources. Sky coordinates, distances, and proper motions of each target are retrieved from the Gaia Data Release 3. Using these data, we compute the positions of each target at the observation epoch and at J2040 by propagating the Gaia coordinates with their proper motions. Figure~\ref{fig:bkg} shows the estimated positions at J2040, with target stars marked by red crosses. For GJ 3473, TOI-1468, and TOI-270, the target star will approach within 2\arcsec\ of a nearby source, which could contaminate future observations.

\subsection{Faint Companions Within 20 AU Separation} \label{sec: close residual}

\begin{figure}
            \centering
            \includegraphics[width=1\linewidth]{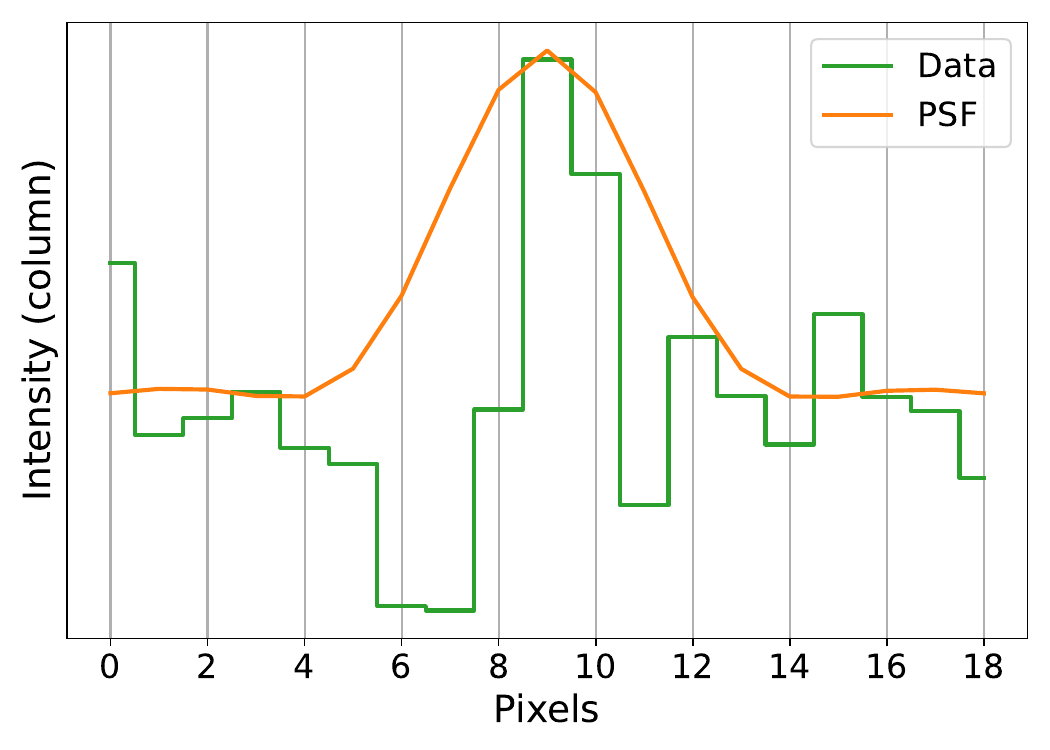}
            \caption{Comparison of observed point-source-like residual to PSF model prediction for HD 260655.
            The green line represents data extracted from the PSF-subtracted image, and the orange line shows the PSF model from interpolating the \texttt{stpsf} grid.
            }
            \label{fig:fm}
\end{figure}

Figure~\ref{fig:image} and Figure~\ref{fig:S/N} reveal a point-source-like residual near the central star HD 260655 (peak S/N 4.86). To determine whether this is an astrophysical signal or artifact, we extract the central column of the residual and compare with the corresponding PSF model obtained by interpolating the \texttt{stpsf} grid. Figure \ref{fig:fm} shows that observed residual is only two pixels wide, which is less than half the expected FWHM of 4.436 pixels. This suggests that this point-source-like residual is most likely an algorithmic artifact rather than real astrophysical signal.

\section{Discussion}

\subsection{MIRI/TSO Mode As A Planet Imager}

Our analysis demonstrates that at 15 $\micron{}$, the sensitivity of JWST allows the detection of Jupiter-mass gas giants at wide orbit. Such time-series imaging data enable us to fill the blank of exoplanet searching at large separations, the region inaccessible by transiting or radial velocity method. Additionally, another 500 hours are approved for the Rocky World directory discretionary time (DDT) program \citep{Redfield2024}. Analyzing method presented in this paper could be applied the DDT program data, enabling the search for wide-orbit gas giants around these targets.

We evaluate the direct imaging sensitivity of MIRI/TSO data by comparing it with published JWST/NIRCam and MIRI results. \citet{bogat2025probing} demonstrated the capability of JWST/NIRCam coronagraphy at 3-5 $\micron{}$ to search for sub-Jupiter mass planets at wide orbits around younger M-dwarfs, with age ranging from 24-440 Myr. They reported sensitivity to Saturn-mass ($\sim0.3\ \mathrm{M_{Jup}}$) planets at Saturn-like projected separations ($\sim$ 9.5 AU). By observing young systems with a coronagraph, the NIRCam survey can directly detect much lower mass planets than those detectable by MIRI/TSO observations of old ($>$5 Gyr) M dwarfs.
The NIRCam survey targeted young M-dwarfs with ages younger than 440 Myr, while the typical age of our targets is $\sim$ 5 Gyr. According to the planet evolutionary models, planets cool and dim as they age. Therefore, beyond the age of 5 Gyr, only the most massive planets remain warm enough to produce sufficient thermal emission detectable by JWST/MIRI. Additionally, their use of coronagraphy effectively suppresses starlight. Another contributor is the fact that MIRI PSF is wider because of the larger diffraction limit. These factors combined contribute to their deeper contrast curves, with a median 5$\sigma$ contrast of $1.5 \times 10^{-5}$ at 1", and a lower planet mass detection limit. While our survey does not reach the same mass sensitivity, it is suited for studying older M-dwarf planetary systems. For planets older than 5 Gyr, the peak of their thermal emission shifts to longer wavelength in the mid-infrared. Therefore, our work at 15 $\micron{}$ provides a valuable window into the population of mature, evolved gas giants at wide separations. 
We further compare our results with coronagraphic MIRI observations. \citet{Matthews2024} observed a K star at 3.6 pc with the F1550C coronagraph at 15.5$\micron{}$, achieving a 5$\sigma$ sensitivity in apparent magnitude of 12.5 mag at 1'' and 13 mag at 3''. This sensitivity is around three magnitudes shallower than ours, likely due to their shorter exposure time of 3,922s. Meanwhile, their contrast in magnitudes is 1.5 magnitudes better than ours at 3", which could be explained by the brighter central K star, which is about 2.1 magnitude at 15.5$\micron{}$. However, their performance degrades more slowly at small separations ($<$2''), demonstrating the coronagraph's effectiveness in suppressing starlight at closer angular separations where our TSO data are limited by residual stellar flux.

Additionally, we compare our results to dedicated MIRI imaging surveys. \citet{BowensRubin2025} reported a 5$\sigma$ contrast of $\sim10^{-3}$ at 1" with F2100W (21 $\micron{}$), corresponding to an sensitivity in apparent magnitude of $\sim12.5$ mag (see their Figure 2c). 
The 3$\sigma$ background-limited region is reached at 2.5", corresponding to an sensitivity in apparent magnitude between 15.5-16 mag. They demonstrated that around Wolf 359, a solar-metallicity M6 M-dwarf with an age between 0.1-1.5 Gyr at 2.4 pc, a planet the same absolute magnitude of Saturn could be detected using MIRI imaging at 21$\micron{}$. Our median sensitivity curve is $\sim$3 magnitude deeper than theirs, which can be partly explained by the wider PSF at longer wavelengths. Nevertheless, since the thermal emission of cold planets peaks at longer wavelengths, their survey achieves lower detectable planet mass limit. In another survey, \citet{Poulsen2024} obtained a 5$\sigma$ contrast of $1.1 \times10^{-2}$ at 0.654" and a median contrast of $6 \times10^{-3}$ beyond 1.26", around a white dwarf at 22.4 pc with F1500W at 15$\micron{}$\footnote{They converted their sensitivity to a planet of 0.5 $\mathrm{M}_{\mathrm{Jup}}$ at 3 Gyr using \texttt{HELIOS} models. While this planet mass limit is lower than ours, it partly comes from the discrepancy between the \texttt{petitCODE} and the \texttt{HELIOS} atmosphere models.}.

Overall, these comparisons demonstrate that the high-contrast imaging performance of MIRI/TSO observations is competitive, and could complement dedicated imaging surveys. The relative performance compared to dedicated surveys depends on several factors including total integration time, reference imaging strategy, and target separation. That being said, achieving comparable performance using free archival data offers approach to probe wide-orbit gas giant populations without additional telescope time.

Furthermore, we checked published Adaptive Optics results for our targets. GJ 3473 \citep{Kemmer2020}, LTT 3780 \citep{Cloutier2020}, TOI-1468 \citep{Chaturvedi2020}, HD 260655 \citep{Luque2022}, and L 98-59 \citep{Kostov2019} have previous AO observations in the Br $\gamma$ band, with a median contrast of $\sim$6.5 mag at 0.5". These results indicate no detectable companions within the field of view for these targets.

One limitation of our work is the lack of planetary evolutionary models with masses below 4 $M_{\mathrm{Jup}}$
at ages older than 5 Gyr. Such models are essential for fully exploiting the sensitivity of time-series observations to old, cold substellar companions, and their development should be prioritized by the community.

\subsection{Multi-Epoch Observations} \label{sec: multiple exposures}

Eight targets in our sample have multiple exposures taken at different epochs. We investigate the impact of combining observations separated in time using TRAPPIST-1 and GJ 3473, which have the longest temporal separations (one year and seven months, respectively). For the remaining targets, after excluding exposures that would introduce apparent residuals (Figure~\ref{fig:exposures}), the usable exposures exhibit positional shifts smaller than 0.1 pixels. Therefore we do not include these targets in this analysis. 

\begin{figure}
            \centering
            \includegraphics[width=1\linewidth]{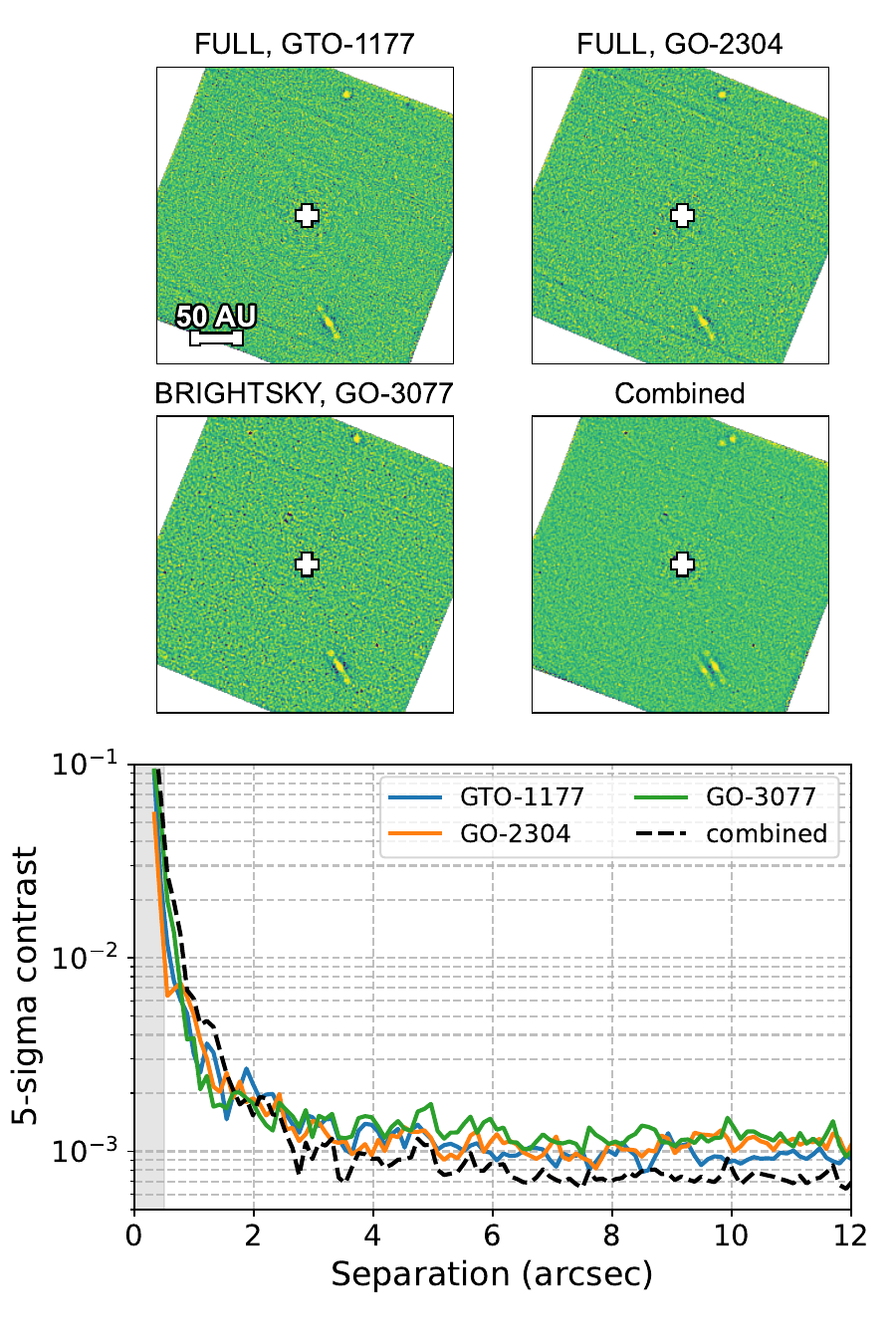}
            \caption{PSF subtraction results for TRAPPIST-1 using individual and combined programs. Upper panel: PSF-subtracted images from each program with observation subarrays labeled. Lower panel: Contrast curves from individual programs, with the combined three-program result shown for comparison.}
            \label{fig:multi}
\end{figure}

TRAPPIST-1 was observed by three separate programs. We run KLIP on both the combined dataset and each program separately. PSF subtraction results and contrast curves are shown in Figure~\ref{fig:multi}. In the combined PSF-subtracted image, two nearby sources show positional offsets relative to TRAPPIST-1 between epochs, indicating they are background sources rather than bound companions.
Combining the three programs improves contrast beyond $2''$ but degrades it at smaller separations. Within $2''$, PSF variations across different observing setups -- particularly subarray choice and epoch -- degrade KLIP subtraction quality. Beyond $2''$, where noise is background-limited, combining programs averages independent noise realizations across datasets and yields improved sensitivity.

\begin{figure}
            \centering
            \includegraphics[width=1\linewidth]{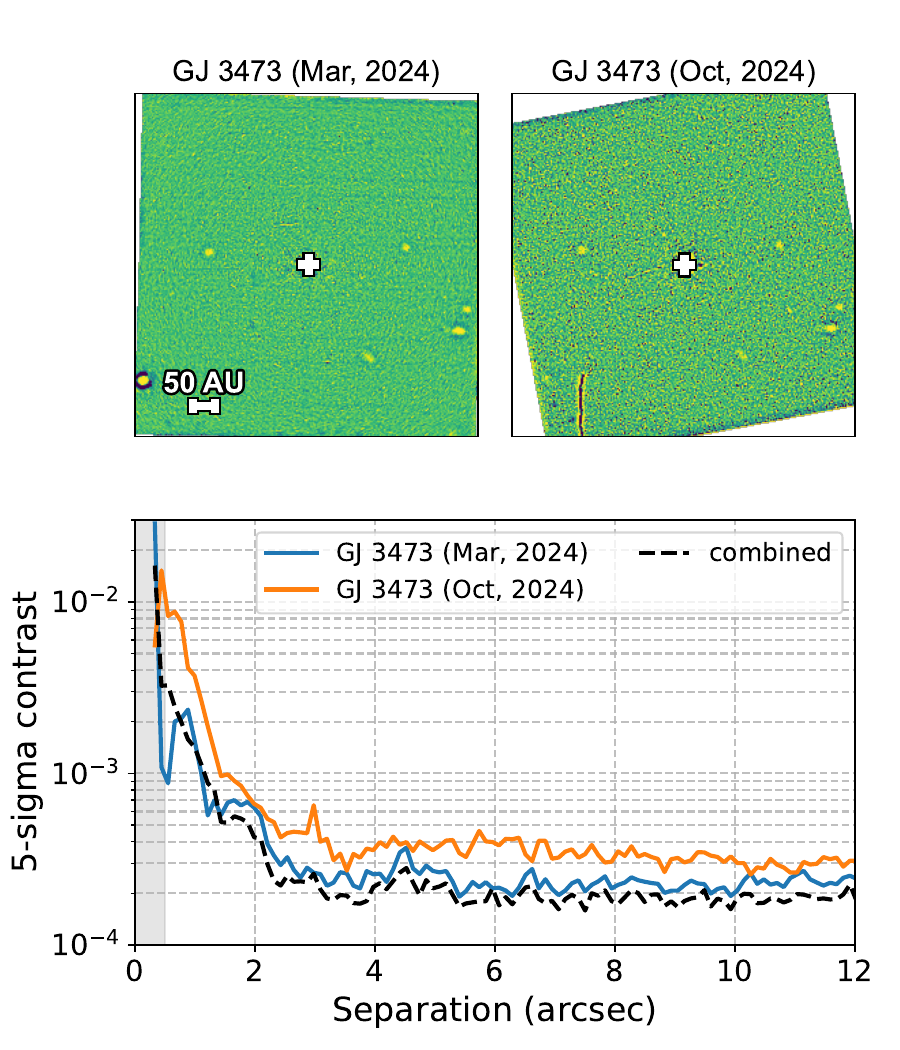}
            \caption{PSF subtraction results for GJ 3473 using individual and combined exposures. Upper panel: PSF-subtracted images from the combined March 12–13 observations (left) and the October 20 observation (right). Lower panel: Contrast curves from individual epochs, with the combined three-exposure result shown for comparison.}
            \label{fig:multi-GJ3473}
\end{figure}

GJ 3473 was observed on 2024 March 12, 13, 30, and October 20. The March 30 observation was excluded from PSF subtraction due to apparent residuals (Figure~\ref{fig:exposures}). We perform KLIP on the combined March 12–13 observations and the October 20 observation separately, with results shown in Figure~\ref{fig:multi-GJ3473}. The combined three-exposure dataset achieves better contrast due to longer integration time. Additionally, nearby sources exhibit 12\% larger major-axis FWHMs in combined images compared to single exposures, consistent with proper motion blur expected for background objects rather than bound companions. This provides a method to distinguish between background sources and bound sources.

We also consider whether co-adding epochs could blur bona fide planetary companions due to orbital motion. For TRAPPIST-1, a bound planet at 10 AU on a face-on orbit would move $\sim$0.4 pixels over the one-year baseline. Since no promising signals were detected in individual epochs, orbital motion corrections were not applied when combining the data.

\subsection{Contrast-Exposure Time Relationship}

We investigate the relationship between contrast and exposure time using 108 images (total exposure time: 56.43 hours) of TRAPPIST-1 from GO-3077. Starting with a single image, we incrementally add images to increase total exposure time. We perform KLIP on each subset using the same parameters as in Section~\ref{sec:image analysis}, and calculate the 5$\sigma$ contrast at separations of 0.5'', 1'', and 5'' for each step. To mitigate the contrast versus time trend fluctuations caused by image order, we randomize the order and repeat this analysis ten times, taking the median contrast-exposure time relation.

\begin{figure}
            \centering
            \includegraphics[width=1\linewidth]{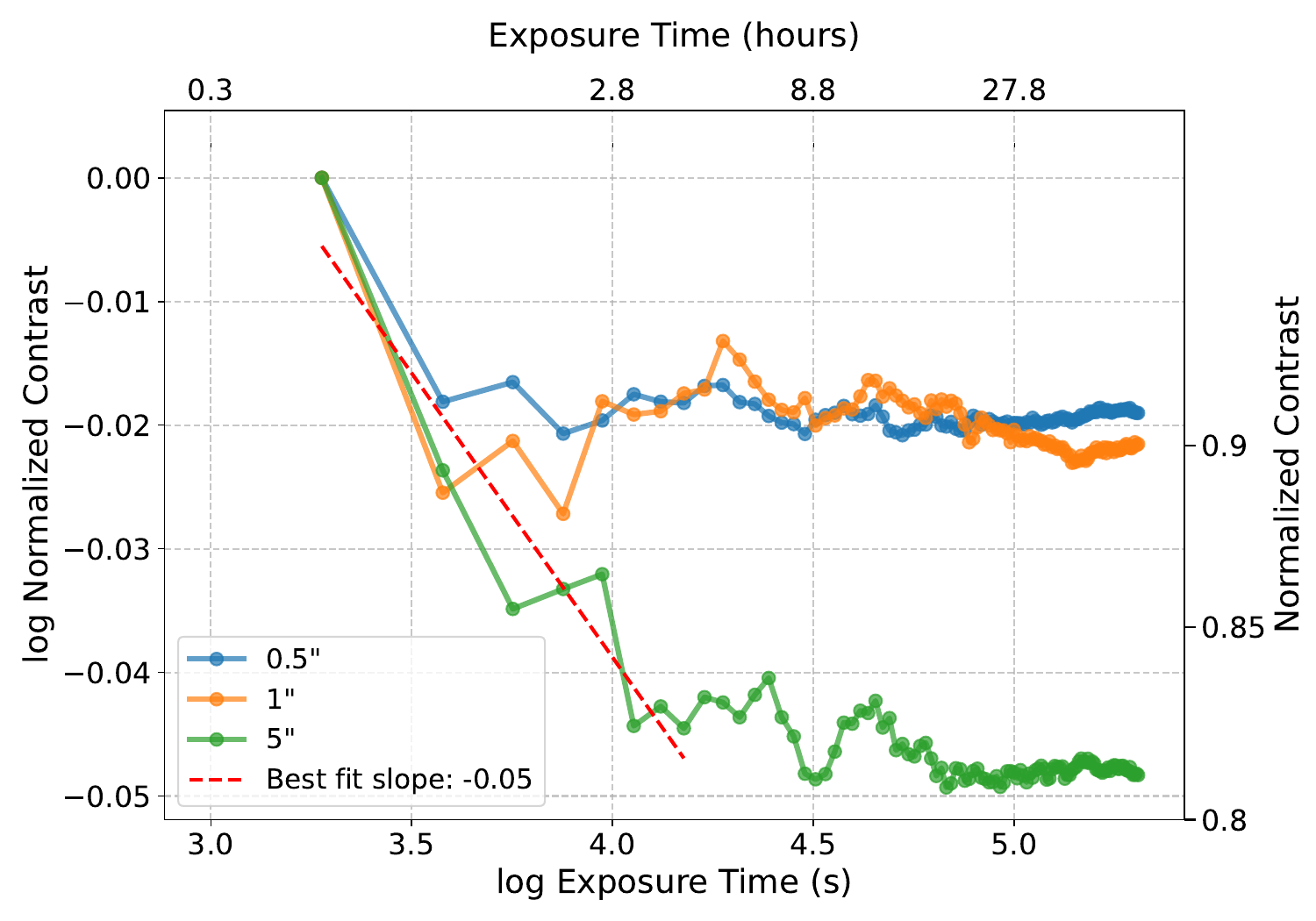}
            \caption{Contrast versus exposure time for TRAPPIST-1 at three angular separations. Blue, orange, and green points represent the median contrast at 0.5", 1", and 5" separations, respectively, computed from ten randomized stacking trials. The red dashed line indicates a linear fit to the initial regime at 5", with a slope of -0.05.Contrast improvement plateaus beyond $\sim$8.8 hours for 5". }
            \label{fig:contrast-time}
\end{figure}

Figure~\ref{fig:contrast-time} shows the resulting relationship in log-log space. To compare the curve shapes, the three curves are normalized by their initial contrast values. At 0.5'' and 1'', contrast remains roughly constant regardless of exposure time. This trend aligns with expectations: at smaller separations, contrast is dominated by PSF subtraction residuals rather than photon noise, so longer exposures provide diminishing returns once systematic errors dominate. Raw contrast at 5'' improves with increasing exposure time before reaching a plateau. The transition occurs at approximately 8.8 hours. The fitted slope in the initial regime at 5'' is -0.05, shallower than the photon noise-limited expectation of -0.5. The deviation likely reflects systematic noise sources that average down more slowly than photon noise.

\subsection{Gas Giants In Multi-Planet Systems}

Our search and its expansion complement previous studies of the relationship between wide-orbit gas giants and close-in terrestrial planets ($\mathrm{P}(\mathrm{GG} \mid \mathrm{E})$). 
\citet{Bryan2024} reported a distinct positive correlation between inner super-Earths and outer gas giants for solar-mass, metal-rich host stars.
While the gas giant frequency is higher around metal-rich M-dwarfs, no significant correlations between super-Earths and gas giants has been established \citep{Bryan2025}.
In our study, we did not detect any wide-orbit companions around our target stars within 50 AU, despite the fact that all of them host transiting terrestrial planets. Due to the limitation of traditional transiting and RV method, previous work has focused on separations between 0.1-10 AU. Our results provide new data points at larger separations ($>$ 10AU), which may assist future study on M-dwarf planetary system.

Our work enables the exploration of the connection between inner planetary system architecture and the occurrence of outer gas giants around M-dwarfs.
\citet{He2023} presented that inner planetary systems around solar-type stars tend to have higher gap complexities ($\mathcal{C}$), a measure of the deviation from uniform spacings, when they also have outer gas giants than when they do not. We examine if similar gap complexity and planet occurrence relation exists in M-dwarfs. 
Using the orbital period data from the NASA Exoplanet Archive, we compute the gap complexity for four systems of our targets with three or more known planets: TOI-270 (0.064), TRAPPIST-1 (0.019), L 98-59 (0.021), and GJ 357 (0.372). For comparison, \citet{He2023} found a median value of $\mathcal{C}\sim$0.06 for systems without outer giants, and $\mathcal{C}>$0.32 for those with outer giant(s). While GJ 357 inner system has relatively high gap complexity comparable to systems with outer giants, we did not detect any companions within 20 AU down to a mass limit of 4 $\mathrm{M}_{\mathrm{Jup}}$. A statistically robust evaluation of this trend rely on improved age constraints, planetary evolutionary models, and a larger target sample.
While our results cannot put new statistical constrains on the theory, we provide a way to study the architecture of M-dwarf systems. The reduction and analysis approach presented in this paper can be applied to more similar time-series data. With large enough datasets, we could put new constrains on the relation between inner and outer planets around M-dwarfs and further understand the gap complexity of its planetary systems.

\section{Summary}

Our findings are:
\begin{enumerate}
    \item
    We performed high-contrast imaging analysis on archival JWST/MIRI F1500W (15\,\micron{}) TSO data of ten intensively studied transiting planet systems and perform direct imaging search for wide-orbit gas giant planets. The ten targets are cross-referenced as PSF reference stars, which supports a successful implementation of reference star differential imaging data reduction. We follow \citet{AdamsRedai2023} and establish a robust framework to optimize PSF subtraction using the KLIP algorithms (Figure~\ref{fig:metrics}). Our analysis deliver the deepest 15\,\micron{} images of our targets (Figures~\ref{fig:image} and \ref{fig:S/N})
    \item
    We achieved a median 5$\sigma$ contrast of $1.5 \times 10^{-3}$ (median sensitivity in apparent magnitude of 15.4 mag) at a separation of 1" and $2.1 \times 10^{-4}$ (17.4 mag) at separations $\gtrsim 3$" (Figure~\ref{fig:contrast}). Assuming 5 Gyr system age, cloud-free, and solar metallicity atmosphere model, we can detect $>$4 $\mathrm{M}_{\mathrm{Jup}}$ planet with a 50\% probability beyond 40 AU in four targets (Figure~\ref{fig:sensitivity}). The detectable mass limit increases with the assumed system age and the sensitivity estimates strongly rely on extrapolating the available planet evolutionary models when system age exceeds 5 Gyr.
    
    \item 
    Beyond 50 AU projected separation, we detected 25 uncataloged sources with high confidence (S/N $> 5$). Table~\ref{tab:bkg} summarizes their relative positions to the central star, detection S/N, brightness in the F1500W band, and their masses based on cooling tracks (assuming that they are substellar companions). If all 25 sources are background stars or galaxies, four of them will be within five arcsec separation to the planet hosts within 15 years and impact future transit observations.
    \item
    Our work enhances the value of MIRI TSO imaging data. The sensitivity in direct detecting planets is competitive against dedicated MIRI imaging surveys. Although the detectable mass limit is higher than NIRCam coronagraphic imaging targeting younger stars, the superior sensitivity to old gas giant planets creates a new way to probe the wide-orbit gas giant planet population. We advocate including the high-contrast imaging analysis as part of the standard routine in processing MIRI TSO imaging observations of transit planet systems.
    \item 
    Our detection probability constraints are limited by the lack of planet evolutionary models at low masses ($<$2 $\mathrm{M}_{\mathrm{Jup}}$) and old ages ($>$5 Gyr). Extending these models to wider parameter space would unlock the full potential of MIRI TSO archives for characterizing the low-mass planet population.
\end{enumerate}

\begin{acknowledgements}
The authors would like to thank Jonas Wehrung-Montpezat for identifying an error in the sensitivity calculation.
Y.L. and Y.Z. acknowledge support from acknowledge support from Heising-Simons Foundation 51 Pegasi b Alumni Faculty Grant (2023-4808 – 51).
We acknowledge support from the Virginia Initiative for Cosmic Origins (VICO) summer program, during which portions of this work were completed.  GJH and YL acknowledge support from National Natural Science Foundation of China general program 12573031 and  grant IS23020 from the Beijing Natural Science Foundation.
This work is based on observations made with the NASA/ESA/CSA James Webb Space Telescope. The data were obtained from the Mikulski Archive for Space Telescopes at the Space Telescope Science Institute, which is operated by the Association of Universities for Research in Astronomy, Inc., under NASA contract NAS 5-03127 for JWST. These observations are associated with programs GTO-1177, GO-2304, GO-3077, and GO-3730.  
\end{acknowledgements}

\facilities{JWST}
\software{\texttt{jwst}\citep{Bushouse2025}, \texttt{pyklip}\citep{Wang2015}, \texttt{photutils}\citep{Bradley2025}, \texttt{astropy}\citep{astropy:2013, astropy:2018, astropy:2022}, \texttt{numpy}\citep{Charles2020}, \texttt{scipy}\citep{2020SciPy-NMeth}, \texttt{matplotlib}\citep{Hunter:2007}, \texttt{lmfit}\citep{Newville2025}}

\bibliographystyle{aasjournalv7}
\bibliography{sample7}

\appendix
\section{Detected background sources}

We list the detected background sources in Table \ref{tab:bkg}.

\begin{longrotatetable}
\begin{deluxetable*}{lrrcrrccccc}
\tablecaption{Background Sources Astrometry And Photometry Properties\label{tab:bkg}}
\tablehead{
\colhead{Source} & \colhead{$\Delta$RA} & \colhead{$\Delta$Dec} & \colhead{Separation} &
\colhead{Mag} & \colhead{S/N} & \colhead{FWHM Major} & \colhead{FWHM Minor} &
\colhead{Mass(1 Gyr)} & \colhead{Mass(5 Gyr)} & \colhead{Mass(10 Gyr)}\\
\colhead{}       & \colhead{(arcsec)}   & \colhead{(arcsec)}    & \colhead{(AU)} & \colhead{}    & \colhead{} &
 \colhead{(pixels)} & \colhead{(pixels)} &
\colhead{($\mathrm{M}_{\mathrm{Jup}}$)} & \colhead{($\mathrm{M}_{\mathrm{Jup}}$)} & \colhead{($\mathrm{M}_{\mathrm{Jup}}$)}
}
\startdata
GJ 3473-1 & 13.49 ± 0.00 & -9.36 ± 0.00 & 448.59 & 13.26 ± 0.01 & 95.0 & 6.1 ± 0.2 & 5.0 ± 0.2 & Extended & -- & -- \\
GJ 3473-2 & 11.06 ± 0.04 & -9.24 ± 0.04 & 393.92 & 16.35 ± 0.10 & 10.3 & 9.7 ± 0.5 & 5.3 ± 0.3 & Extended & -- & -- \\
GJ 3473-3 & -4.75 ± 0.03 & -7.44 ± 0.03 & 241.09 & 16.26 ± 0.08 & 13.3 & 8.7 ± 0.4 & 5.1 ± 0.2 & Extended & -- & -- \\
GJ 3473-4 & -11.99 ± 0.02 & -5.25 ± 0.01 & 357.63 & 15.42 ± 0.04 & 24.9 & 8.1 ± 0.2 & 4.7 ± 0.1 & Extended & -- & -- \\
GJ 3473-5 & -12.68 ± 0.03 & -3.55 ± 0.03 & 359.71 & 16.23 ± 0.11 & 9.4 & 5.5 ± 0.2 & 4.1 ± 0.2 & Extended & -- & -- \\
GJ 3473-6 & 8.16 ± 0.02 & 1.06 ± 0.02 & 224.71 & 15.98 ± 0.07 & 15.1 & 5.9 ± 0.2 & 4.9 ± 0.2 & Extended & -- & -- \\
GJ 3473-7 & -7.75 ± 0.02 & 1.43 ± 0.02 & 215.42 & 16.29 ± 0.08 & 13.2 & 4.8 ± 0.2 & 4.3 ± 0.2 & 2.3 & 6.6 & 10.1 \\
LHS 1478-1 & -1.75 ± 0.01 & -5.16 ± 0.01 & 99.37 & 14.23 ± 0.03 & 37.0 & 7.7 ± 0.1 & 7.0 ± 0.1 & Extended & -- & -- \\
LHS 1478-2 & 3.30 ± 0.00 & 10.25 ± 0.00 & 196.15 & 13.59 ± 0.01 & 147.1 & 4.4 ± 0.0 & 4.1 ± 0.0 & 10.2 & 29.2 & 40.9 \\
TOI-270-1 & 2.15 ± 0.01 & -13.39 ± 0.01 & 304.96 & 13.34 ± 0.04 & 23.9 & 4.9 ± 0.1 & 4.5 ± 0.1 & Extended & -- & -- \\
TOI-270-3 & 3.55 ± 0.02 & -7.24 ± 0.02 & 181.30 & 16.37 ± 0.06 & 16.1 & 6.2 ± 0.3 & 5.2 ± 0.2 & Extended & -- & -- \\
TOI-270-4 & 0.60 ± 0.03 & -6.05 ± 0.03 & 136.74 & 16.45 ± 0.09 & 11.5 & 7.4 ± 0.4 & 5.3 ± 0.3 & Extended & -- & -- \\
TOI-270-5 & 4.54 ± 0.02 & 0.19 ± 0.02 & 102.20 & 15.77 ± 0.05 & 19.5 & 6.6 ± 0.2 & 5.0 ± 0.1 & Extended & -- & -- \\
TOI-270-6 & 8.45 ± 0.08 & 12.14 ± 0.09 & 332.41 & 17.04 ± 0.36 & 2.9 & 9.4 ± 0.5 & 5.6 ± 0.3 & Extended & -- & -- \\
TOI-270-7 & 7.56 ± 0.04 & 6.25 ± 0.04 & 220.66 & 17.27 ± 0.14 & 7.4 & 5.6 ± 0.4 & 5.3 ± 0.4 & Extended & -- & -- \\
TOI-1468-1 & 13.58 ± 0.04 & -10.40 ± 0.04 & 423.02 & 15.76 ± 0.18 & 5.9 & 6.2 ± 0.7 & 6.2 ± 0.7 & 2.9 & 8.1 & 12.4 \\
TOI-1468-2 & -0.00 ± 0.00 & -7.63 ± 0.00 & 188.65 & 12.58 ± 0.00 & 232.6 & 5.8 ± 0.1 & 4.5 ± 0.0 & Extended & -- & -- \\
TOI-1468-3 & -1.44 ± 0.01 & -4.14 ± 0.01 & 108.50 & 16.14 ± 0.05 & 20.6 & 5.2 ± 0.2 & 4.8 ± 0.2 & Extended & -- & -- \\
TOI-1468-4 & 5.57 ± 0.00 & -3.82 ± 0.00 & 166.92 & 13.73 ± 0.02 & 64.6 & 6.4 ± 0.0 & 4.8 ± 0.0 & Extended & -- & -- \\
TOI-1468-5 & 7.55 ± 0.09 & 7.11 ± 0.08 & 256.54 & 16.48 ± 0.59 & 1.8 & 4.7 ± 0.2 & 4.5 ± 0.2 & 2.0-2.1 & 4.7 & 7.4 \\
TOI-1468-6 & -6.72 ± 0.00 & 8.65 ± 0.00 & 270.81 & 12.52 ± 0.01 & 156.2 & 5.4 ± 0.0 & 5.1 ± 0.0 & Extended & -- & -- \\
TOI-1468-7 & 9.08 ± 0.00 & 11.25 ± 0.00 & 357.41 & 12.59 ± 0.00 & 387.5 & 4.4 ± 0.0 & 4.2 ± 0.0 & 44.0 & 74.5 & 75.9 \\
TOI-1468-8 & -4.60 ± 0.07 & 2.67 ± 0.06 & 131.39 & 16.60 ± 0.22 & 4.7 & 8.0 ± 0.4 & 5.7 ± 0.3 & Extended & -- & -- \\
TRAPPIST-1-1 & -1.98 ± 0.02 & -9.95 ± 0.02 & 126.48 & 14.91 ± 0.06 & 18.2 & 14.0 ± 0.3 & 4.6 ± 0.1 & Background & -- & -- \\
TRAPPIST-1-2 & -3.66 ± 0.02 & 11.36 ± 0.02 & 148.89 & 15.31 ± 0.09 & 11.8 & 4.6 ± 0.2 & 4.1 ± 0.1 & Background & -- & -- \\
\enddata
\tablecomments{Sources detected on the PSF subtracted images of all targets. We list each source with its position relative to the target star, apparent magnitude from forward modeling, PSF fitting S/N, fitted major and minor axis FWHM, and estimated mass under the assumption that they are planets at age 1 Gyr, 5 Gyr, and 10Gyr. Starred FWHM values indicate uncertain fits for sources located near the image edge. For sources whose magnitudes fall outside the coverage of both the \texttt{BEX-petitCODE} grids and the \texttt{ATMO-CEQ} grids, we give the upper and lower mass limit given by these two models. Sources labeled "Extended" have major axis FWHM exceeding 3$\sigma$ of the point source FWHM (4.436 pixels), indicating they are likely background galaxies or other extended objects rather than point-source companions. }
\end{deluxetable*}
\end{longrotatetable}

\end{document}